\pdfoutput=1
\documentclass{emulateapj}
\shorttitle{Evolution of the ICM}
\shortauthors{O'Hara et al.}

\newcommand{\Chandra}{{\it Chandra}}
\newcommand{\ROSAT}{{\it ROSAT}}

\newcommand{\XMM}{XMM-{\it Newton}}
\def\Micm{$M_{g}$}

\def\Micmfive{$M_{g, 500}$}
\def\Micmtfive{$M_{g, 2500}$}
\def\L2500{$L_{2500}$}

\def\M2500{$M_{2500}$}
\def\rtfive{$r_{2500}$}
\def\rd{$r_\delta$}
\def\rfive{$r_{500}$}

\def\AI{$A_{I}$}
\def\RI{$R_{I}$}
\def\LX{$L_{\rm X}$}
\def\LXCS{$L_{\rm XCS}$}
\def\OmegaM{$\Omega_M$}
\def\OmegaL{$\Omega_\Lambda$}
\def\LXCSfive{$L_{\rm XCS,500}$}
\def\LXCStfive{$L_{\rm XCS,2500}$}
\def\LXfive{$L_{\rm X,500}$}
\def\LXtfive{$L_{\rm X,2500}$}
\def\ficm{$f_{g}$}
\def\sint{$\sigma_{\rm int}$}

\def\Rsix{$R_{6\times10^{-14}}$}
\def\Rthree{$R_{3\times10^{-14}}$}
\def\Rone{$R_{1.5\times10^{-13}}$}
\def\Tx{$T_{\rm X}$}
\def\Txcs{$T_{\rm XCS}$}
\def\NH{$N_{\rm H}$}

\def\Io{$I_0$}
\def\OTI{$\mathcal{O}$--$T_{\rm X}$--$I_0$}

\def\Ez{$E(z)$}

\def\myputfigure#1#2#3#4#5%
{\vskip#5pt\makebox[0pt]{\hskip#2in
\includegraphics[width=#3\textwidth]{#1}}\vskip#4pt\hfill}
\newenvironment{inlinefigure}{
\def\@captype{figure}
\noindent\begin{minipage}{0.999\linewidth}\begin{center}}
{\end{center}\end{minipage}\smallskip}

\begin{document}

\submitted{Submitted to ApJ August 26, 2007}

\title{Evolution of the Intracluster Medium Between \lowercase{$0.2 < z < 1.3$} in a {\it Chandra} Sample of 70 Galaxy Clusters}

\author{Timothy B. O'Hara,\altaffilmark{1,2} Joseph J. Mohr,\altaffilmark{1,2} \& Alastair J.~R. Sanderson\altaffilmark{3}}

\altaffiltext{1}{Department of Physics, University of Illinois,
1110 West Green St, Urbana, IL 61801; tbohara@astro.uiuc.edu, jmohr@uiuc.edu}
\altaffiltext{2}{Department of Astronomy, University of Illinois,
1002 West Green St, Urbana, IL 61801}
\altaffiltext{3}{School of Physics and Astronomy, University of Birmingham, Edgbaston, Birmingham B15 2TT, UK}

\begin{abstract}

We study the evolution of the intracluster medium (ICM) with a uniformly analyzed sample of 70 galaxy clusters spanning $0.18 < z < 1.24$ and observed with \Chandra. We find that X-ray luminosity and ICM mass at a fixed temperature evolve with redshift in a manner inconsistent with either the standard self-similar model of cluster formation or a model that assumes no evolution of cluster structure. Both luminosity and ICM mass evolve more slowly than the self-similar prediction, i.e., clusters have lower luminosity and ICM mass at fixed emission-weighted temperature than expected at higher redshifts. We find that evolution in these two observables can be modeled by a simple evolution in the cluster gas mass fraction, evolving as $(1+z)^{-0.39\pm0.13}$ when measured using core-subtracted observables. Excluding cluster cores from measurements results in evolution more consistent with the self-similar model than when the entire cluster is used, indicating that the fraction of clusters with cool cores increases with time, or that cool cores become more developed over time in those clusters that have them; this is supported by direct study of the redshift dependence of central surface brightness, which increases in scatter and magnitude at low redshift. We find that isophotal size--temperature relations evolve differently according to which isophote is used, indicating that the central and outer regions of cluster ICM evolve differently. We show that  constraints on the evolution of the gas fraction and isophotal size--temperature relations constraints can be combined to measure cluster distances, and thus to constrain cosmological parameters in a way complementary to other techniques. Scatter in scaling relations is considerably reduced by using either core-subtracted quantities or three-parameter relations including the central surface brightness; in addition, there are indications that scatter decreases at higher redshift, suggesting that merging is not the dominant source of cluster structural variation. Our results provide constraints for simulations attempting to model cluster physics, indicate some difficulties for cosmological studies that assume constant cluster gas fractions, and point toward other potentially more robust  uses of clusters for cosmological applications.

\end{abstract}


%%%%%%%%%%%%%%%%%%%%%%%%%%%%%%%%%%%%%
%%%%%%%%%%%%%%%%%%%%%%%%%%%%%%%%%%%%%
\section{Introduction}
\label{sec:intro}

Scaling relations among bulk properties of galaxy clusters provide a powerful means to test models of the large-scale structure and evolution of the universe. These correlations among properties such as X-ray luminosity, intracluster medium (ICM) mass, mean ICM temperature, and cluster virial mass reflect gravitational and non-gravitational processes involved in the formation of structure in an expanding universe. Scaling relations also provide the means to readily estimate masses of clusters from much more easily measured properties such as luminosity, an essential component of cosmological studies that use X-ray observations to determine the redshift evolution of the cluster mass function.

Simple models of cluster formation via gravitational collapse predict particular forms for the redshift evolution of cluster scaling relations \citep[][]{kaiser86}. Adding additional cluster physics such as radiative cooling of the ICM, and energy injection by active galactic nuclei (AGN), supernovae, and star formation,  modifies these predictions \citep[e.g.,][]{cavaliere98,ettori04b,muanwong06,kay07}. Observational studies of scaling relation evolution are required to properly constrain models of cluster evolution and to understand the effects of non-gravitational processes on the scaling relations that will be used to study cosmology. X-ray studies of the ICM are complementary to studies of the evolution of the cluster galaxy population \citep[e.g.,][]{depropris99,lin06}, helping to constrain the overall evolution of cluster baryons and their distribution in various forms within clusters.

 Several studies of X-ray scaling relation evolution  have been carried out in recent years \citep[e.g.,][]{vikhlinin02,ettori04a,kotov05,maughan06,morandi07,branchesi07b}, but no clear consensus has emerged. In this paper we will address scaling relation evolution using a systematic analysis of a	 \Chandra\ sample of 70 clusters covering $0.18 < z < 1.24$, the largest sample yet used for this purpose. 

Our study addresses two difficulties which may affect scaling evolution measurements. The first arises from the fact that radiative cooling of the ICM leads to the development of cool, dense (and hence very luminous) cores in many clusters; these relatively small cores bias cluster measurements such as X-ray temperature and luminosity to an extent that they are not representative of the overall cluster structure. This introduces significant scatter into scaling relations; indeed,  there is evidence that cool core clusters, which are traditionally regarded as ``relaxed," actually exhibit greater structural variation than non-cool core clusters, which are often thought to have recently undergone major mergers \citep{ohara06}. Studies of scaling relations commonly attempt to ``correct" for the impact of cool cores on cluster properties by one of several methods, such as simply leaving clusters with evidence for strong cool cores out of the sample \citep[e.g.,][]{arnaud99}, or excising central regions within a fixed metric radius \citep[e.g.,][]{morandi07} or a fraction of the virial radius \citep[e.g.,][]{maughan07}, and perhaps ``correcting" measured luminosity by some factor determined from a model of the cluster surface brightness distribution \citep[e.g.,][]{vikhlinin02}. In this paper we measure temperatures with and without cores defined as fractions of the virial radius, and we also measure luminosities with and without the same core. By using relations both with and without core subtracted quantities, we can examine the effects that core development has on cluster scaling relation slopes and evolution.

The other issue usually faced by scaling relation studies is the use of scaling relation slopes and normalizations from low-redshift studies carried out with different instruments. The relatively small fields of view of \Chandra\ and XMM-{\it Newton} make measurements of local samples quite challenging with those instruments; hence, studies using older X-ray instruments are used as references for $z=0$ relations. Unfortunately, differences in spectral and imaging results among X-ray instruments are well established, making such approaches subject to instrument-related systematics; indeed, even the same instrument has produced results differing by the author, as calibrations change and varying reduction and analysis methods are adopted. By using a large sample (70 clusters), we can avoid the use of outside references for scaling relation parameters or the direct inclusion of data from other samples, in favor of a single, homogeneously analyzed sample. While this approach is not entirely new---for example, \citet{branchesi07b} studied evolution using their own 17 cluster sample both with and without the inclusion of data from other studies; and \citet{morandi07} studied a homogeneously reduced 24 cluster sample---the size of our sample leads to significantly smaller uncertainties on scaling relation parameters than have otherwise been obtained.

In \S~\ref{sec:theory} we provide a brief overview of scaling relations and their predicted evolution, and in \S~\ref{sec:data} we explain our data reduction and measurement procedures. We test for scaling relation evolution with respect to expectations from the self-similar theory and from a scenario of no evolution in cluster parameters in \S\S~\ref{sec:scaling_SS} and \ref{sec:scaling_noev}, respectively, and provide an explanation for observed evolution in scaling relations via a simple evolution in the gas mass fraction \S~\ref{sec:ficm}. In \S~\ref{sec:size} we examine the evolution of isophotal size, and discuss the implications for studying cosmology using size measurements, and in \S~\ref{sec:scatter} we discuss the effectiveness of two different methods of reducing the scatter in measured scaling relations. In \S~\ref{sec:discuss} we compare our results to previous observations and simulation results, and discuss some implications of our findings. Finally, we list our conclusions in \S~\ref{sec:concl}.

We adopt the {\it WMAP} + LRG $\Lambda$CDM  cosmology from \citet{spergel07}, which combines the third year {\it WMAP} data with results from the SDSS luminous red galaxy survey \citep{eisenstein05} to give $H_0 = 70.9$ ~km~s$^{-1}$~Mpc$^{-1}$,  $\Omega_M=0.266$, and $\Omega_\Lambda = 0.734$.
All uncertainties are 68\% confidence, unless specified otherwise.

%%%%%%%%%%%%%%%%%%%%%%%%%%%%%%%%%%%%%
%%%%%%%%%%%%%%%%%%%%%%%%%%%%%%%%%%%%%
\section{Scaling Relation Background}
\label{sec:theory}

The self-similar model \citep[e.g.,][]{kaiser86} describes formation of clusters via gravitational collapse of overdense regions in an expanding universe. In this model the ICM is heated by this gravitational collapse and the resulting shock heating, but no so-called non-gravitational heating is assumed. As a result, clusters scale self-similarly, i.e., they scale only because of changes in their physical size at fixed mass due to density variation as the universe expands. With the assumptions of spherical symmetry, hydrostatic equilibrium, a constant gas fraction, and X-ray emission dominated by thermal bremsstrahlung, this leads to X-ray luminosity \LX\ and ICM mass \Micm\ scaling with ICM temperature \Tx\ and redshift as

\begin{equation}
L_{\rm X} \propto T_{\rm X}^2 E(z),
\end{equation}

\begin{equation}
M_{g} \propto T_{\rm X}^{3/2} E(z)^{-1},
\end{equation}

\noindent
where $E(z)$ is the ratio of the Hubble parameter at redshift $z$ to its present value. In a flat cosmology with matter density $\Omega_{\rm m}$, $E(z)$ has the form:

\begin{equation}
E(z) = H(z)/H_0 = \left[ \Omega_{\rm m}(1+z)^3 + 1- \Omega_{\rm m} \right]^{1/2}.
\end{equation}

Predicting scaling laws for the isophotal size (i.e., the physical size of the region corresponding to the angular size of a particular X-ray isophote; see \S~\ref{sec:imaging}) requires additional assumptions about the ICM mass distribution. If the  ICM distribution scaled self-similarly with mass, then isophotal size scales as

\begin{equation}
R_{I} \propto T_{\rm X}^{2/3},
\end{equation}

\noindent
with no redshift evolution if the ICM density falls off as $r^{-2}$ outside the core \citep[i.e., $\beta=\frac{2}{3}$;][]{mohr00}.  

Observational studies have found that scaling relations for all three of these observables (\LX, \Micm, and \RI) in fact have a stronger dependence on temperature than predicted by self-similar models \citep[e.g.,][]{edge91,markevitch98b,mohr97a,mohr99}. Explanations for this and other evidence of non-gravitational processes, such as the presence of entropy ramps in the central regions of clusters \citep[e.g.,][]{ponman03},  typically involve additional  energy injection by active galactic nuclei (AGN), supernovae, and star formation \citep[e.g.,][]{bialek01,bower01,borgani02,mccarthy04,kay07}; radiative cooling of the ICM, which leads to the formation of cool, dense cores in many clusters; and non-radiative cooling \citep[e.g.,][]{bryan00}. 

It is important to note that there are multiple ways to define radii for measuring cluster parameters, which result in different predicted redshift evolution for scaling relations. The expressions given above are correct for observables (\LX\ and \Micm) measured within regions corresponding to fixed overdensities relative to the critical density. This is appropriate for our strategy in this paper, in which we choose to measure cluster properties within virial regions defined by local relations, and then test for consistency with the evolution scenarios described below. Another commonly used form for the redshift evolution of scaling relations \citep[e.g.,][]{ettori04a,branchesi07b,morandi07} uses densities defined from assumptions of virial equilibrium in a spherical collapse model. These densities have their own redshift evolution, leading to additional factors in the scaling relation evolution equations. In either case, it is common to parametrize additional redshift evolution beyond the self-similar predictions in terms of a simple power law with redshift, i.e., proportional to $(1+z)$ raised to some power.
 
In this paper we discuss two models for cluster evolution. The first is ``self-similar evolution," in which cluster observables scale as would be expected given purely gravitational influence as discussed above, i.e., $L_{\rm X} \propto E(z)$ and $M_{g} \propto E(z)^{-1}$. The other is what we will refer to as ``no evolution," meaning that cluster parameters, including  virial radii, do not scale at all as the universe expands.

 \begin{inlinefigure}
   {\myputfigure{Tx_red}{1.5}{1.8}{-300}{-70}}
   \figcaption{\label{fig:Txred} 
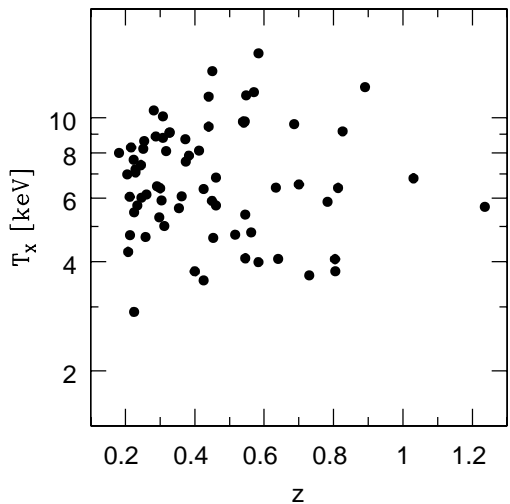
Measured emission-weighted mean temperature \Tx\ plotted versus redshift for the clusters in our sample.
}
\end{inlinefigure}

%%%%%%%%%%%%%%%%%%%%%%%%%%%%%%%%%%%%%
%%%%%%%%%%%%%%%%%%%%%%%%%%%%%%%%%%%%%
\section{Data Reduction}
\label{sec:data}

\subsection{The Cluster Sample}

The data are drawn from the \Chandra\ archive. The lower redshift limit of $z\sim0.2$ reflects the difficulty in measuring cluster parameters out to at least \rtfive\ for clusters closer than this, given the small \Chandra\ field of view. The cluster sample is listed in Table~\ref{tab:basicinfo},  with the ID number of the \Chandra\ observation used for each cluster.

Having been largely developed through cluster selection in archival {\it Einstein} IPC and {\it ROSAT} PSPC observations, our sample is essentially X-ray flux limited. However, as the sample is not derived from a single homogeneous survey at a fixed flux threshold, it might be expected to include systematically more luminous (i.e., more massive) systems at high redshifts. In Figure~\ref{fig:Txred}  we plot the emission-weighted mean temperatures for our sample (measured as described in \S \ref{sec:spec} below) versus redshift. Our sample spans a consistent range of \Tx over the full redshift range.

\subsection{X-ray Data Reduction}

The data reduction is carried out using the standard \Chandra\ analysis software {\sc ciao}, version 3.3, with {\sc caldb} version 3.2.1,  and the spectral fitting package {\sc xspec}, version 11.3.1. We generate new level 2 events files from the level 1 files obtained from the \Chandra\ archive, so that all observations are reduced in a uniform manner. The following reduction procedure is applied to each cluster.

Light curves are extracted for back-illuminated chips 5 and 7 individually, and for front-illuminated chips 0--3 and 6 combined. Light curves are extracted and binned in time using the recommended criteria for each chip.\footnote{http://cxc.harvard.edu/ciao/} Flares are excluded using the {\sc ciao} task ``{\sc lc\_clean}" based on the median value of the light curve. The exposure times after filtering are given in Table~\ref{tab:basicinfo}.

Cosmic ray events are identified with the {\sc ciao} tool ``{\sc acis\_run\_hotpix}". A new level 1 events file is then generated using the latest gain file, and charge transfer inefficiency (CTI) and time-dependent gain variation corrections are applied as appropriate. Standard bad columns and hot pixels are excluded. Events with ASCA  grades of 0, 2, 3, 4, and 6 are used. A level 2 events file is then created from the filtered level 1 events file. Where the observation was made in very faint (VF) mode, we carry out the extra background event flagging that this enables.

We attempt to use background data from the actual data sets, extracting the background from regions well away from target cluster or other emission. For some clusters, however, emission fills most of the detector, and in these cases we extract the background spectrum from the Markevitch blank-sky data.\footnote{http://cxc.harvard.edu/contrib/maxim/acisbg/} To account for small differences in the particle background between these statistical backgrounds and each individual observation, the blank-sky sets' exposure times are scaled by the ratio of counts in the 7--12 keV energy band in the data and blank-sky observations. Before using either background method point sources are identified by the iterative method described in \citet{sanderson05} and checked by visual inspection, and then excluded. Even when emission-free regions are available, if the spectral fit is worse with the local background than with the blank-sky background, we use the latter. In total, we use the blank-sky backgrounds for 41 of the 70 clusters in our sample.

\subsection{Spectral Fitting}
\label{sec:spec}

\begin{figure*}[t]
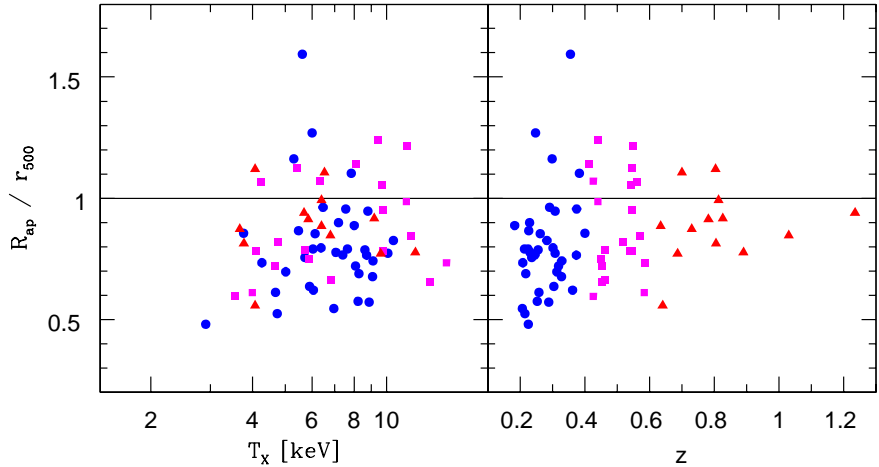

   {\myputfigure{aperture}{7}{.8}{-270}{-70}}
   \figcaption{\label{fig:aperture} 
The ratio of our spectral extraction radius to \rfive\ (defined as described in \S\ref{sec:imaging}) for each cluster, plotted versus the measured non-core subtracted temperature ({\it left}) and versus redshift ({\it right}). Markers correspond to $z<0.4$ (circles), $0.4<z<0.6$ (squares), and $z>0.6$ (triangles).
}
\end{figure*}

Cluster spectra are extracted in regions  with maximum radius chosen by eye to be   where the cluster emission merges into the background; the center coordinates and radii of our extraction regions are given in Table~\ref{tab:basicinfo}. Choosing apertures based on the X-ray surface brightness  distribution might result in smaller apertures relative to the physical size of clusters that are cooler or lie at higher redshifts, and thus will tend to have observations with fewer total counts. However, Figure~\ref{fig:aperture}, which plots the ratio of the spectral extraction radius to \rfive\ for each cluster versus cluster mean temperature ({\it left}) and redshift ({\it right}), suggests that this is not the case. The mean ratio of aperture radius to \rfive\ is $0.84 \pm 0.20$ (RMS), with no apparent temperature or redshift dependence.

 We generate weighted response matrix files (RMFs) using the {\sc ciao} tool {\sc mkacisrmf} when the data allow; otherwise we use the older tool {\sc mkrmf}.

We fit to the cluster spectra a single-temperature APEC  model with a component for galactic absorption. We use fit \NH\ values when they are reasonable (i.e., within a few standard deviations of the galactic value), and not pegged to zero; otherwise, we fix \NH\ to the galactic value \citep{dickey90}. In total, we fit \NH\ for 18 of the 70 clusters. We generally extract spectra in energy bands of 0.7--9 keV for ACIS-I, and 0.5--9 keV for ACIS-S. In a few cases we use an upper limit of 7 keV when there is clearly spurious, non-background emission above this value; in no case does this change the measured temperature at greater than the 1--2\% level. We use the Cash statistic \citep{cash79}, which is preferable to $\chi^2$ when the number of photons is low. In our sample the use of the Cash statistic generally results in a best-fit temperature that is a few percent higher than that measured with $\chi^2$.

We measure the core subtracted temperature \Txcs\ by extracting spectra with the same maximum radius as described above, but excluding the inner 0.2\rfive; the core subtracted temperature and the 0.2\rfive\ exclusion radius are measured iteratively until convergence. (Our definition of \rfive\ is given in \S~\ref{sec:imaging}.) For two clusters, ZwCl 1356+6245 and CLG J0647+7015, the iteration does not converge to a reasonable value when the core is excluded, and so we do not measure core subtracted quantities for those two clusters.

Our measured values for the temperature of the entire cluster, and for \Txcs\ measured assuming self-similar evolution and assuming no evolution, are given in Table~\ref{tab:basicinfo}.

\subsection{Comparison with Published Temperatures}

\tabletypesize{\scriptsize}
\begin{deluxetable*}{ccccccccc}
\tablewidth{0pt}
\tablecaption{Comparison of Temperature Measurements}
\tablehead{
& & & & \multicolumn{2}{c}{Balestra et al. (2007)} & & \multicolumn{2}{c}{Maughan et al. (2007)} \cr 
\cline{5-6} \cline{8-9}
Cluster & $z$ & Our \Tx\ (keV) & & They (keV) & We\tablenotemark{a} (keV) & & They (keV) & We\tablenotemark{a} (keV) 
}
\startdata
MS 0451.6-0305 & 0.54 & $9.8\pm0.8$ & & $8.2^{+0.4}_{-0.3}$ & $10.5 \pm 0.7 $ & & $6.7^{+0.6}_{-0.5}$ & $8.1\pm0.4$ \\ [.3em]
ClG J1149+2223 & 0.54 & $9.8 \pm0.8$ & & $12.9^{+1.2}_{-1.0}$ & $9.9^{+1.0}_{-0.8} $ & & $8.4^{+0.9}_{-0.7}$ & $8.7^{+0.9}_{-0.6}$ \\ [.3em]
ClG J1120+2326 & 0.56 & $4.2^{+0.6}_{-0.3}$ & & $5.2\pm0.5$ & $4.4^{+0.4}_{-0.3} $ & & $3.8^{+0.4}_{-0.3}$ & $3.2\pm0.3$ \\ [.3em]
ClG J1113-2615 & 0.73 & $3.7^{+0.6}_{-0.5}$ & & $5.7^{+0.9}_{-0.6}$ & $5.0^{+0.9}_{-0.8} $ & & $3.8^{+0.9}_{-0.7}$ & $3.1\pm0.4$ \\ [.3em]
RX J1317+2911 & 0.81 & $3.8^{+1.7}_{-0.9}$ & & $4.5^{+1.4}_{-1.0}$ & $4.4^{+0.9}_{-1.0} $ & & $2.0^{+0.7}_{-0.5}$ & $3.3^{+1.5}_{-0.8}$ \\ [-.7em]
\enddata
\tablenotetext{a}{Our measurement of the cluster temperature using the same aperture and similar methods as the literature sources; see text.}
\label{tab:Txcomp}
\end{deluxetable*}

 \begin{figure*}[t]
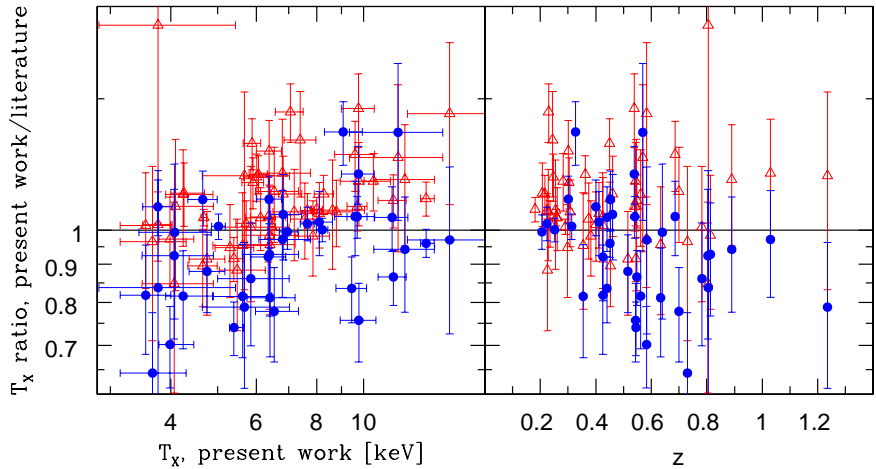

   {\myputfigure{Bothcomp}{7}{.8}{-270}{-70}}
   \figcaption{\label{fig:Txcomp} 
The ratio of our measured cluster temperatures to published temperatures (vertical axis), plotted versus our temperature ({\it left}) and versus redshift ({\it right}). Published temperatures are from \citet{maughan07} (triangles) and \citet{balestra07} (circles).
}
\end{figure*}

Though calibrations continue to improve, measurements of the same cluster by different instruments, and by different methods with the same instruments, lead to temperature measurements that differ. To check the accuracy of our own temperature measurements, we  compare our  values to those obtained in two other recent \Chandra\ studies. 

\citet{balestra07} (hereafter Ba07) studied 56 clusters over a redshift and temperature range similar to our own; our samples have 38 clusters in common. Our data reduction and spectral fitting processes differ from theirs in several small ways: Ba07 use local backgrounds exclusively, while we, as described above, use  blank-sky backgrounds when local backgrounds are not possible or give worse spectral fits; they always fix the value of \NH\ to galactic, while we allow it to float when the value obtained thereby is reasonable; they use a spectral extraction band of 0.6--8 keV, versus our 0.5-- or 0.7--9 keV; and they include a spectral component to compensate for Ir-M edge residuals, a correction that has been taken into account in the more recent calibration files which we use. Because clusters are not isothermal, the emission-weighted mean temperature is affected by the choice of energy band. Most importantly, Ba07 use spectral extraction regions determined via a method intended  to maximize the $S/N$, which results in the use of extraction radii up to a factor of two smaller than ours. Their resulting extraction regions have a clear redshift trend, with radii as small as $\sim$0.3 \rfive\ at high redshift.

\citet{maughan07} (hereafter Ma07) measured temperatures for 115 clusters, of which 53 are in common with our sample. Differences between our analyses include their use of a 0.6--9.0 keV spectral fitting band; their fixing \NH\ to the galactic values; and their use of blank-sky backgrounds in some cases where we use local backgrounds, plus an additional soft X-ray background component. Ma07 also use a different method for determining the spectral extraction region, measuring all spectra out to a radius of \rfive\ as determined from an iterative procedure using a mass--$Y_{\rm X}$ relation, where $Y_{\rm X}$ is the product of the temperature and gas mass \citep{kravtsov06}.

To examine the difference between our temperatures and those of these two studies, we compare the error-weighted ratio of our temperatures to theirs. Overall, our temperatures are {\it lower} than those of Ba07 by a weighted average of $(3\pm1)$\%, and {\it higher} than those of Ma07 by $(6\pm1)$\%. To examine whether we can reproduce their values, we remeasured the temperatures of five clusters using methods similar to those of Ba07 and Ma07; i.e., we used their reported aperture radii, spectral extraction bands, and spectral models. We fixed \NH\ in all cases for this comparison, but did not change our choice of background strategies. As can be seen in Table~\ref{tab:Txcomp}, these changes resulted in generally higher temperatures when using the methods closer to those of Ba07, and generally lower temperatures when using methods closer to those of Ma07, thus at least partially explaining the sources of systematic differences between our measurements and those of these two papers. Note that this does not mean that our temperatures necessarily came to agree more closely with theirs; for MS 0451.6-0305, for example, our initial temperature was higher than that of Ba07, and these changes resulted in an even higher temperature.

The overall hotter temperatures that we measure relative to Ma07 may be attributable to variations in ICM temperature with radius. As shown in \S\ref{sec:spec}, our spectral extraction radii average ($0.84\pm0.20$)\rfive, while Ma07 uses uniform radii of \rfive. The ICM temperature generally decreases with radius at these radii \citep[e.g.,][]{vikhlinin05,pratt07}, and so we would expect our measured temperatures to be systematically slightly higher than those of Ma07.

However, the differences between our temperature measurements and those of the other two studies are not uniform; there are dependences on temperature and, for Ba07, redshift. The left panel of Figure~\ref{fig:Txcomp} shows the ratio of our temperatures to the literature values versus our temperature. In the case of the Ma07 comparison, the ratio is clearly greater at higher  temperatures; a one-dimensional least-squares fit of a straight line shows that the ratio increases as $(0.14^{+0.12}_{-0.19}) \log T_{\rm X}$ for Ba07, and as $(0.22^{+0.05}_{-0.09}) \log T_{\rm X}$ for Ma07. 
The latter trend may again result from Ma07's choice of \rfive\ as an extraction radius; extraction regions of hotter clusters may include more background-dominated area, leading to temperature systematics as parts of the spectrum are deweighted by background noise.

The right panel of Figure~\ref{fig:Txcomp} shows the dependence of temperature ratio on redshift. There is no evidence for a redshift dependence when comparing to Ma07; the ratio varies as $(0.02\pm 0.09)z$. For Ba07, however, the ratio varies as $(-0.31^{+0.09}_{-0.08})z$, showing a clear negative dependence on redshift. This is almost certainly a result of Ba07's use of extraction regions that feature a trend toward smaller fractions of the virial radius at higher redshift.

The differences between our measured temperatures and those from the literature underscore the difficulties inherent in comparing cluster parameters measured using differing instruments, instrumental calibrations, and methods. This calls into question the reliability of results obtained from directly combining data from multiple studies \citep[e.g.,][]{branchesi07b}, and suggests that caution should be taken when comparing more processed results, such as the low-redshift slopes and normalizations often combined with new measurements of higher-redshift clusters to test for scaling relation evolution \citep[e.g.,][]{vikhlinin02,ettori04a,kotov05}.

\subsection{Imaging Analysis}
\label{sec:imaging}

We extract X-ray images and use the spectral fit to obtain the conversion factor from counts to physical units in the rest frame 0.5--2 keV band. Because the flattening of statistical backgrounds using the exposure map generated for a particular observation results in a spatially inhomogeneous background image, we fit a flat background to the regions outside of the cluster emission using the same technique used to determine the surface brightness profile, described below. The results of this fitting are checked by examining radial brightness profiles and via simple comparison of total counts in regions well outside of cluster emission. 

As our observations do not in general contain enough photons to do a deprojection analysis, particularly at high redshift, we  fit the  standard spherical $\beta$ model \citep{cavaliere78} to the cluster emission:

\begin{equation}
\label{eq:betamod}
I\left(r\right) = I_{0} \left[1+\left(\frac{r}{R_c}\right)^2\right]^{-3\beta+1/2},
\end{equation}

\noindent
with central brightness $I_0$, core radius $R_c$, and power-law index $\beta$. In what are traditionally considered ``cool core" clusters, i.e., where there is a central emission excess due to the formation of a cool dense core, we fit a double $\beta$ model \citep{ikebe96,ikebe99,mohr99} with both components having the same center coordinates and index $\beta$, so that the total surface brightness is the sum of the two, i.e.,

\begin{equation}
\label{eq:doublebetamod}
I\left(r\right) = \displaystyle\sum_{i=1} ^2  I_{0,i} \left[1+\left(\frac{r}{R_{c,i}}\right)^2\right]^{-3\beta+1/2}.
\end{equation}

We fit these surface brightness profiles to the two-dimensional surface brightness images, and find the best fit and one $\sigma$ confidence intervals for each parameter using the Cash statistic. In a few cases cases (A521, A1682, and A2744) there are prominent clumps or subclusters separate from the main body of the cluster, which are masked out before fitting.  In two cases we fix the value for $\beta$: A521, a multiply-merging cluster \citep{ferrari03}, for which we find a somewhat stable value of $\beta=0.75$, which we adopt over the values of $\beta > 3$ which are found by a full gridding analysis; and ClG J1056-0337, a merging system \citep{jee05} for which we find only very high values of $\beta$, leading us to adopt the canonical $\beta=0.67$. In both cases we then measure 2 $\sigma$ uncertainties in the other fit parameters. The $\beta$ model parameter fit results are listed in Table~\ref{tab:betainfo}. The second, bright central component is, where used for a given cluster, listed as the second brightness and core radius components $I_2$ and $R_{c,2}$.

We measure several different cluster observables, each of which---X-ray luminosity, ICM mass, isophotal size, and mean ICM temperature---derives from the underlying cluster structure in a different way; by studying the evolution of multiple observables, we are examining the evolution of the ICM in multiple ways. Luminosity and ICM mass are measured within two different virial radii \rd, which permits us to examine evolution on different scales within a cluster. We determine \rfive\ and \rtfive\ from the cluster temperature using $M_\delta$--\Tx\ relations determined by \citet{arnaud05} using \XMM\ observations of local galaxy clusters. We use their relations for clusters with $T_{\rm X} > 3.5$ keV:

\begin{equation}
r_{500} = 1.129\left(\frac{T_{\rm X}}{5~{\rm keV}}\right)^{0.497}E(z)^{-1}~{\rm Mpc},
\end{equation}
\begin{equation}
r_{2500} = 0.501\left(\frac{T_{\rm X}}{5~{\rm keV}}\right)^{0.503}E(z)^{-1}~{\rm Mpc}.
\end{equation}

\noindent
Note that by using virial radii obtained in this manner, we are implicitly testing the evolution of these local mass--temperature relations along with our other observables. That is, our ``self-similar evolution" scenario includes evolution of the \rd--\Tx\ relations as written above, and the ``no evolution" scenario includes no evolution (i.e., no \Ez\ factor) in the \rd--\Tx\ relations.

We measure the projected X-ray luminosity \LX\ in the rest frame 0.5--2 keV band from the images described above, within radii of \rfive\ and \rtfive; we also measure core subtracted luminosities \LXCS\ by excising the projected luminosity from the central 0.2\rfive. Luminosity measurements are centered on the cluster brightness peak, with the exception of A521, where we use the peak brightness of the main cluster, not the brighter infalling subcluster to the north of the cluster center \citep[see, e.g.,][]{ferrari06}; and ClG J1056-0337, where we use the western brightness peak, which has been identified as the ``central" mass peak via weak lensing \citep[e.g.,][]{jee05}. Given the small field of view of Chandra, the virial radii \rd\ often extend beyond the image boundary; furthermore, some observations are not deep enough that there is signal measurable out to a given  \rd. We thus establish for each cluster a maximum radius from the brightness peak at which either the detector edge is reached or the $S/N$ falls close to unity; in a few cases the maximum radius is determined by the presence of other structure, as in the cases of ACO 2246 and ClG J1701+6414, which lie a small angular distance from one another in the same observation. Then, if the radius \rd\ exceeds this established maximum radius for a given luminosity measurement, we do not carry out that measurement on that particular cluster; this is reflected in Tables~\ref{tab:SSinfo} and \ref{tab:noEinfo}, where luminosity measurements are not given in many cases. We include in the luminosity uncertainties contributions from the temperature used in calculating \rd, as well as a uniform 10\% background uncertainty.

The X-ray luminosity within a given radius can be modeled analytically by an integral of the ICM density profile and X-ray emissivity out to that radius. We can therefore use a measurement of the actual luminosity together with the measured $\beta$ model parameters and the cluster temperature to find the central ICM density, and  hence ICM mass via an integral of the density function to a given radius of interest; for details see \citet[][]{mohr99}. We estimate uncertainties on \Micm\ by including the statistical uncertainties on the $\beta$ model fit; a uniform 10\% background uncertainty in the luminosity measurement; and temperature uncertainties in \rd. The ICM mass measurement is not subject to the same maximum radius restriction as luminosity, as the luminosity within any given radius can be used to measure the central density; while larger luminosity measurement radii are of course preferable, it is not necessary to measure the flux out to a given \rd\ for an ICM mass measurement within that radius. Note that we do not similarly use the $\beta$ model to extrapolate luminosity measurements out to a radius of interest; this is because we prefer to directly use projected luminosities without assumptions as to the structure of the cluster, but ICM mass cannot similarly be measured without such assumptions.

We measure the isophotal size \RI\ of a cluster by measuring the area \AI\ enclosed by an isophote $I$, and finding the effective radius given by $A_I = \pi R_I^2$. For these measurements we use images that have been adaptively smoothed using the {\sc ciao} task {\tt csmooth}. We include the 10\% background uncertainty in the \RI\ uncertainties by remeasuring at isophotes increased and decreased by the background uncertainty. In the  0.5--2 keV band we are using here, the conversion from X-ray counts to physical units varies slowly with cluster temperature, so we do not include temperature uncertainties in the isophotal size. We measure \RI\ at three different isophotes, $1.5\times10^{-13}$, $6\times10^{-14}$, and $3\times10^{-14}$ erg s$^{-1}$ cm$^{-2}$ arcmin$^{-2}$ (in the rest frame 0.5--2 keV imaging band), which, like using both \rfive\ and \rtfive\ for the luminosity and ICM mass, lets us study evolution of \RI\ on different scales within a cluster. Clusters can ``fall off" an isophotal size--temperature scaling relation when the isophote used approaches the peak surface brightness of the cluster; we therefore exclude clusters when their measured isophotal size is less than 0.2\rfive, our adopted core exclusion region.

 \begin{figure*}[h]
   {\myputfigure{lumpanel}{7}{.7}{-160}{-40}}
   \figcaption{\label{fig:lumpanel} 
Projected X-ray luminosity within \rtfive\ (left) and \rfive (right), with non-core subtracted (top) and core subtracted (bottom) quantities, plotted versus temperature. These quantities are measured assuming {\it self-similar evolution}. Luminosity values are scaled to $z=0$ using the best-fit redshift scaling for each relation, and the best-fit slope is plotted for each relation. Markers correspond to $z<0.4$ (circles), $0.4<z<0.6$ (squares), and $z>0.6$ (triangles).
}
   {\myputfigure{masspanel}{7}{.7}{-160}{-40}}
   \figcaption{\label{fig:masspanel} 
Same as Figure~\ref{fig:lumpanel}, but for \Micm--\Tx\ relations.
}
\end{figure*}

\subsection{Fitting Procedures}
\label{sec:fitting}

For a given relation involving an observable $\mathcal{O}$, we fit the form 

\begin{equation}
\mathcal{O} = A  \left( \frac{T_{\rm X}}{6~{\rm keV}} \right)^\alpha (1+z)^\gamma,
\end{equation}

\noindent
or, in log space,

\begin{equation}
\label{eq:logrelation}
\log \mathcal{O} = \log A + \alpha\log\left( \frac{T_{\rm X}}{6~{\rm keV}} \right) + \gamma \log (1+z).
\end{equation}

\indent
That is, we fit a power-law temperature dependence $\alpha$, power-law redshift dependence $\gamma$, and $A$, the normalization at zero redshift and temperature 6 keV.

In this paper we use unweighted orthogonal fits, meaning that we minimize the sum of the square of the point-line orthogonal distances, i.e., the sum

\begin{equation}
\label{eq:orthfit}
\displaystyle\sum_i \left\{ \frac{\log \mathcal{O}_i - \left[\log A+ \alpha \log (T_{{\rm X},i}/6)+\gamma\log(1+z_i)\right]}{\left(1+\alpha^2\right)^{1/2}} \right\}^2 .
\end{equation}

\noindent
Note that the form for redshift evolution assumed here  is evolution of the normalization only, and so there is no factor of $\gamma$ in the denominator. We determine 1 $\sigma$ uncertainties  via bootstrap sampling; the best-fit value given in this paper is the mode of a histogram constructed from the bootstrapping results, and the 1 $\sigma$ confidence interval is constructed in the usual manner so as to contain 68.3\% of the counts around this mode. We also give here the RMS scatter in the vertical dimension (e.g., in \LX\ in the \LX--\Tx\ relation) for the best-fit parameters; this one-dimensional scatter is a more intuitively understandable quantity than the orthogonal scatter, as it reflects the scatter in an observable (\LX, \Micm, \RI) at a given temperature. We refer to this as the intrinsic scatter (\sint), as the measurement uncertainties are generally much smaller than the total scatter in these relations \citep[e.g.,][]{ohara06}.

The question of which fitting method is ``best" is still open, and rests to a large extent on whether one property (such as \Tx) is considered more fundamental than the other (such as \LX); this often seems implicit in discussions of \LX--\Tx, \Micm--\Tx, and other relations, and would imply that a one-dimensional least-squares fit, with temperature (the lowest scatter mass estimator) as the independent variable, might be appropriate. But if both observables are considered to be linked via another property of the system (such as cluster mass), then a  orthogonal minimization fit, which treats both variables equally, may be  more appropriate; we take this view, and so adopt orthogonal fitting in this paper. 

Fits of mock scaling relations using both the orthogonal fitting method and an ordinary least-squares (OLS) fit support this decision. A difficulty that arises in such tests is that assumptions must be made regarding the  scatter in mock relations; e.g., if only scatter in the $y$ direction is generated, then an OLS fit will doubtless give better results than an orthogonal fit. 
For example, \citet{lopes06} make the claim that orthogonal regression produces more accurate measurements of scaling relation slopes than the bisector method (discussed below),  based on their own tests using mock data sets; however, as these data sets were generated using orthogonal scatter, such a result is entirely expected. Because of the difficulty in defining ``correct" scatter, we test scenarios in which only scatter in the $y$ direction is used, and in which equal scatter in both the $x$ and $y$ directions is used. That is, we generate a random $x$ value, use an assumed scaling relation to find $y$, and then shift the values using normal random deviates in the $y$ direction only, or in both the $x$ and $y$ directions. Note that using equal $x$ and $y$ scatter is {\it not} the same as using orthogonal scatter, and so an orthogonal relation should not be {\it a priori} assumed to give the correct result in such a case.  In our testing we do not assume measurement uncertainties, but fit an intrinsic scatter in the $y$ direction in the OLS fits so that the reduced $\chi^2$ value is equal to unity. Again, in real scaling relations the scatter is generally dominated by intrinsic scatter, so this is a reasonable approach.

The results of our tests clearly indicate that the OLS method is a less robust approach than the orthogonal method. For example, when using only $y$ direction scatter of 0.05 (i.e., the random deviates have a standard deviation in $\log_{10}$ space of 0.05), the orthogonal method gives a result that is 2\% ($\sim$1$\sigma$) high while the OLS method gives the correct result; but when using equal scatter of 0.05 in $x$ and $y$, the orthogonal method gives the correct slope, while the OLS method gives a result that is $\sim$10\% ($\sim$2$\sigma$) too low. The results get worse for OLS more rapidly than for orthogonal fitting; e.g., scatter of 0.15 in $y$ only gives an orthogonal slope that is 16\% ($\sim$2.5$\sigma$) high, but scatter of 0.10 in both dimensions gives an OLS result that is 51\% ($\sim$8.5$\sigma$) too low. The results are very similar when true orthogonal scatter is used, rather than random, but on average equal, scatter in each dimension.

Again, the exact origin of scaling relation scatter is unknown, so it is difficult to declare a ``correct" way of testing fitting methods. There is undoubtedly some measurement scatter, however, and so scaling relations certainly have at least some scatter in both dimensions. For this reason, as well as the physical arguments given above, we adopt the orthogonal fit as our chosen method for this paper.

Besides orthogonal fitting, another  approach that treats the two variables equally is the bisector method, in which OLS fits are done with each of the two variables as independent and dependent (i.e., $y$ as a function of $x$, and $x$ as a function of $y$), and the final result bisects the two individual fits. This is not appropriate for our work, because we fit observables as a function of both temperature and redshift, and it is unclear how the bisector method can be extended into three dimensions. Orthogonal fitting is clearly defined in any number of dimensions; i.e., it seeks the shortest point-line distance in two dimensions, the shortest point-plane distance in three dimensions, and so forth. Also, each individual OLS fit in the bisector method is subject to the great dependence on the form of scatter as discussed above, and so the bisector method's utility for studying scaling relations is likewise questionable.

%%%%%%%%%%%%%%%%%%%%%%%%%%%%%%%%%%%%%
%%%%%%%%%%%%%%%%%%%%%%%%%%%%%%%%%%%%%
\section{Tests of the Self-Similar Evolution Scenario}
\label{sec:scaling_SS}

We now examine the evolution of scaling relations while assuming self-similar evolution, as discussed in the introduction. That is, we assume that \rd\ scales as \Ez$^{-1}$ when measuring \LX\ and \Micm, and when determining the core subtraction radius for \Txcs\ and \LXCS. Our values for \LX\ and \Micm, measured using the non-core subtracted temperature, are given in Table \ref{tab:SSinfo}. We then test whether scaling relations evolve in a manner consistent with self-similar evolution.

\tabletypesize{\scriptsize}
\begin{deluxetable*}{cccccc}
\tablewidth{0pt}
\tablecaption{Fit Parameters Assuming Self-Similar Evolution}
\tablehead{
\multicolumn{6}{c}{Core Subtracted Relations} \cr
& & & & Diff. from 0 & \\
Relation & $\alpha$ & $A$\tablenotemark{a} & $\gamma$ & (\%)\tablenotemark{b} & \sint\tablenotemark{c}   
}
\startdata
%$^{+}_{-}$
\LXCStfive--\Txcs &  $2.00^{+0.23}_{-0.19}$ & $2.00^{+0.35}_{-0.30}$E44& $-0.86^{+0.38}_{-0.36}$ & $95-$ & $0.28\pm0.05$  \\ [0.3em]
% $0.12\pm0.02$
\LXCSfive--\Txcs & $2.26^{+0.29}_{-0.33}$ & $3.02^{+1.35}_{-1.24}$E44 & $-1.28^{+1.28}_{-0.86}$ & $69-$ & $0.21^{+0.08}_{-0.07}$  \\ [0.3em]
% $0.09\pm0.03$
\Micmtfive--\Txcs & $1.63^{+0.09}_{-0.08}$ & $2.57^{+0.18}_{-0.17}$E13  & $-0.35^{+0.20}_{-0.15}$ & $90-$ & $0.00^{+0.05}_{-0.00}$  \\ [0.3em]
% $0.00^{+0.02}_{-0.00}$
\Micmfive--\Txcs & $1.56\pm0.10$ & $7.94^{+0.06}_{-0.05}$E13 & $-0.24^{+0.20}_{-0.18}$ & $74-$ & $0.09 \pm 0.04$  \\ [0.3em]
% $0.04 \pm 0.02$

\tableline 
\multicolumn{6}{c}{Non-Core Subtracted Relations} \cr
& & & & Diff. from 0 &\\
Relation & $\alpha$ & $A$\tablenotemark{a} & $\gamma$ & (\%)\tablenotemark{b} & \sint\tablenotemark{c}   \\
\tableline \tableline 
%$^{+}_{-}$
\LXtfive--\Tx & $2.75^{+0.29}_{-0.26}$  & $3.24^{+0.75}_{-0.61}$E44  &  $-1.50^{+0.42}_{-0.49}$ & 99.4$-$ & $0.60^{+0.08}_{-0.09}$ \\ [0.3em]
% $0.26\pm0.04$
\LXfive--\Tx & $2.35^{+0.33}_{-0.24}$ & $6.03^{+3.75}_{-2.22}$E44  & $-1.90^{+1.17}_{-1.11}$ & $90-$ & $0.39^{+0.12}_{-0.10}$ \\ [0.3em]
% $0.17^{+0.05}_{-0.04}$
\Micmtfive--\Tx & $1.82\pm0.08$ & $2.69^{+0.19}_{-0.18}$E13 & $-0.55^{+0.17}_{-0.15}$ & $99.4-$ & $0.14\pm0.02$  \\ [0.3em]
%$0.06\pm0.01$
\Micmfive--\Tx & $1.74\pm0.09$ & $8.32^{+0.59}_{-0.56}$E13 & $-0.45^{+0.18}_{-0.16}$  & $98-$ & $0.13\pm0.02$ 
%$0.06\pm0.01$ 

\enddata
\tablenotetext{a}{In units of $L_\odot$ for \LX--\Tx\ relations, $M_\odot$ for \Micm--\Tx\ relations.}
\tablenotetext{b}{Significance level at which $\gamma$ differs from zero, as determined by bootstrap sampling and refitting; the sign indicates whether $\gamma$ is positive ($+$) or negative ($-$).}
\tablenotetext{c}{Intrinsic scatter in \LX\ or \Micm\ at fixed temperature, expressed in base $e$.}
\label{tab:SSfits}
\end{deluxetable*}

 \begin{figure*}[t]
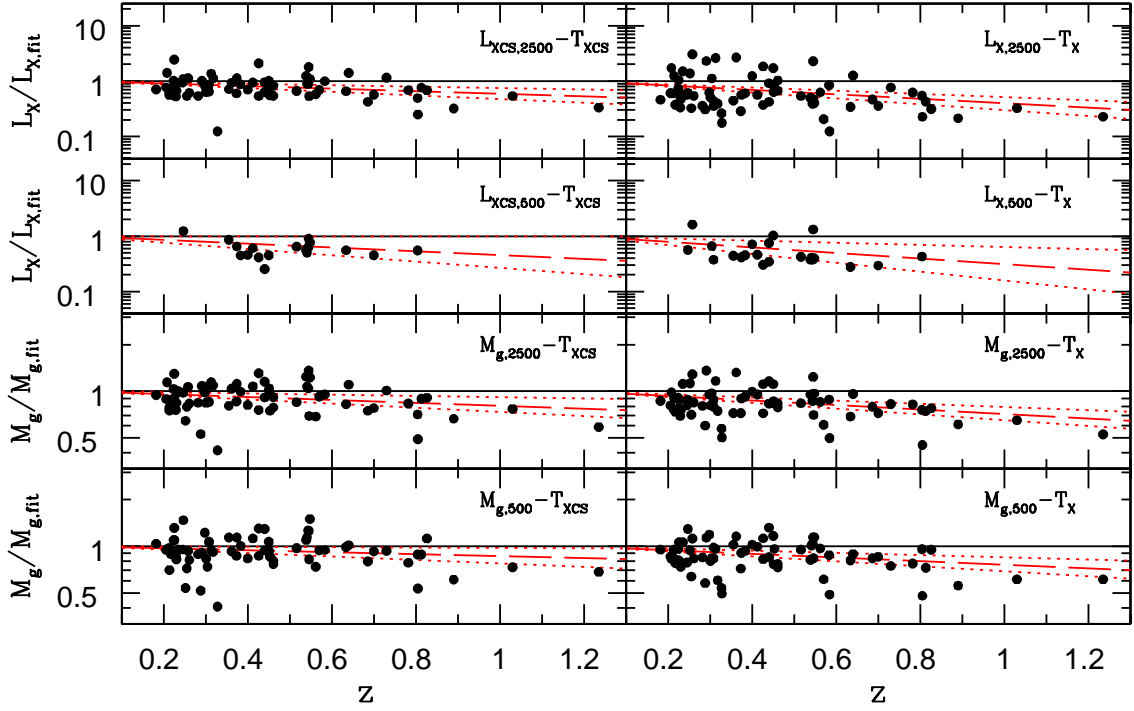

   {\myputfigure{all_evol_ydir}{7}{1}{-300}{-60}}
   \figcaption{\label{fig:evol_E} 
Ratio of measured observable (luminosity or ICM mass) to the best-fit  observable--temperature scaling relation, plotted versus redshift. These measurements assume {\it self-similar evolution}. The horizontal line ($\mathcal{O}/\mathcal{O_{\rm fit}}=1$)  corresponds to no evolution beyond the assumed self-similar evolution, i.e., $\gamma=0$ in our notation. The dashed and dotted lines correspond to the best fit and $1 \sigma$ boundaries on $\gamma$ for each relation.
}
\end{figure*}

\subsection{Scaling Relations}

The \LX--\Tx\ and \Micm--\Tx\ scaling relations are plotted in Figures \ref{fig:lumpanel} and \ref{fig:masspanel}, respectively. In these figures the observables are scaled to $z=0$ using the best-fit scaling relations. One qualitative feature of note is that the scatter is clearly smaller in the \LXCS--\Tx\ relations than in their non-core subtracted counterparts; a similar, though smaller effect is visible in the ICM mass relations. This difference in scatter arises from biases in both temperature and the other observable in each relation induced as a result of cool core-related phenomena \citep[e.g.,][]{fabian94c,markevitch98b,ohara06}. Another interesting feature is the shallowness of the \LXCStfive--\Txcs\ relation compared to the non-core subtracted \LXtfive--\Tx\ relation. Best-fit scaling relation parameters are given in Table~\ref{tab:SSfits}.

Studies of scaling relation evolution commonly fix the slopes to values measured from local samples, and fit only for an evolution factor. Because we are fitting all parameters simultaneously, we need to compare our measured slopes to those of local samples. Our \LXtfive--\Tx\ relation and \LXfive--\Tx\ relation have  slopes of $2.75^{+0.28}_{-0.26}$ and $2.35^{+0.33}_{-0.24}$, respectively, which are significantly higher than the self-similar expectation $\alpha=2$, as has been generally observed in low-redshift samples \citep[e.g.,][]{markevitch98b}; note that using luminosities from a fixed energy band as done here (rest frame 0.5--2 keV) gives a somewhat lower slope than the more commonly used bolometric luminosities, as shown by, e.g., \citet{zhang07}.  For the \Micmtfive--\Tx\ relation we find $\alpha=1.82\pm0.08$, in good agreement with $\alpha = 1.91\pm0.16$ found by \citet{ettori02} using {\it BeppoSAX} data and a bisector fit; for the \Micmfive--\Tx\ relation we find $\alpha = 1.74\pm0.09$, in fair agreement with $\alpha = 1.98 \pm 0.11$ measured by \citet{mohr99} using {\it ROSAT} PSPC images and a mixture of {\it Einstein}, {\it Ginga}, and {\it ASCA} temperatures, with an unweighted orthogonal fit. Both of these are significantly higher than the self-similar expectation $\alpha=1.5$. 

In all cases, the scaling relations with core subtracted quantities have shallower slopes than the standard relations. Remarkably, the core subtracted relations have slopes consistent with the self-similar expectation to within $1 \sigma$, the sole exception being \Micmtfive--\Txcs, which is consistent to within $2 \sigma$.

\subsection{Evolution with Redshift}
\label{sec:Eevol}

Figure~\ref{fig:evol_E} shows the ratio of observables (\LX\ and \Micm) to the self-similar expectation, plotted versus redshift. That is, the vertical axis is the ratio of the observed value to the self-similar prediction using the appropriate fit in Table \ref{tab:SSfits} and the cluster temperature and redshift. The horizontal line in each plot therefore marks the self-similar expectation.
Plotting in this way shows deviations from the self-similar redshift evolution prediction as a redshift dependence of the ratio $\mathcal{O}_i / \mathcal{O}_{\rm fit} =0$;  we also plot the best-fit value of $\gamma$ for each relation, showing how the normalization of each scaling relation in fact evolves.

For each scaling relation, Table~\ref{tab:SSfits} includes the percent significance by which each relation differs from zero, i.e., the significance of its deviation from the self-similar prediction. Because the distributions of $\gamma$ are not in general normal, this significance is determined using binned data to measure the probability density at $\gamma=0$, and integrating to the same probability density on the other side of the peak value. Because we use binned data to estimate this parameter,  it can be determined most precisely when $\gamma$ is significantly different from zero; hence, we quote only at 1\% precision for values less than 99\%. 

All luminosity-- and ICM mass--temperature scaling relations  have $\gamma<0$ at greater than the 1$\sigma$ level.  There is clearly an overall tendency for relations to evolve more slowly than expected from the self-similar prediction, i.e., $\gamma < 0$. We can combine multiple probabilities by assuming independence of the scaling relations; though all of the measured properties are of course linked to some extent by their dependence on the underlying ICM structure, X-ray luminosity and ICM mass depend on that structure in very different ways, and the two virial radii which we use probe two rather different regions of the cluster (i.e., \rfive\ comes close to looking at the cluster as a whole, while \rtfive\ measures a much smaller fraction that is more dependent on core structure and evolution). Combining the results for  all four core subtracted relations by multiplying the given probabilities of consistency with zero gives a combined probability of $<0.1\%$ that all four relations are consistent with the self-similar evolution scenario, ruling out pure self-similar evolution at greater than 3$\sigma$ confidence. The same relations with non-core subtracted quantities have an even smaller probability (i.e., $\ll0.1\%$) of consistency with zero. 

We draw your attention to the $z>0.8$ clusters in our sample because of the special leverage they have on our evolution results.  Examination of Figure~\ref{fig:evol_E} suggests no qualitative difference in the high-redshift population when compared to lower-redshift clusters. For these clusters to bias our results toward more negative evolution, it would require systematically selecting {\it underluminous} clusters, which is the opposite of what is expected. 

The relations involving core subtracted quantities have more positive evolution than those involving non-core subtracted quantities. This could indicate a decrease in clusters with cool cores at higher redshifts, which is expected in the scenario wherein clusters form cool cores over time in the absence of major merging events. The evolution of the cool core fraction remains relatively unexplored; \citet{bauer05} found no evolution in the cool core fraction up to $z\sim0.4$ using spatially resolved spectral analysis, but such an analysis is difficult to carry out at higher redshifts. \citet{vikhlinin06} used a measurement of the ``cuspiness" of the surface brightness distribution to count cool cores in a sample of clusters at $z>0.5$, and found a fourfold decrease in the cool core fraction from z=0 to z=0.5, which might support the concept of cool cores indicating a ``relaxed cluster" that has not undergone recent major mergers. This concept is being increasingly challenged, however, by results from simulations that ascribe the presence or lack of a cool core to aspects of cluster formation history such as preheating \citep{mccarthy04} or early major mergers \citep{burns07}, and observational evidence that cool core and non-cool core cluster populations differ in characteristics beyond their morphological state \citep{ohara06}. \citet{burns07} specifically studied the redshift evolution of the cool core fraction, and find no change  in the fraction up to $z\sim1$ in simulations that successfully reproduce other aspects of cluster and core structure. Our results here may support the classical notion of cool cores evolving over time, in support of the \citet{vikhlinin06} results. Alternatively, a constant cool core fraction could still produce an apparent negative evolution in scaling relation normalization simply because cool cores in those clusters that do have them will tend to grow over time; such a result was reported in simulations by \citet{kay07}. We further discuss possible evolution in scatter in \S~\ref{sec:scatter}.

\subsection{Summary of Self-Similar Evolution Results}

X-ray luminosity and ICM mass at fixed temperature evolve more slowly than expected from the self-similar evolution model. This conclusion is supported by significant ($>1\sigma$) negative evolutions in all \LX-- and \Micm--\Tx\ scaling relations, and by combined constraints using multiple core subtracted or non-core subtracted relations that rule out self-similar evolution at $>99.9\%$ confidence. The less negative evolution of the core subtracted relations suggests that the cool core fraction decreases with redshift, that cool cores grow over time, or a combination of the two.

%%%%%%%%%%%%%%%%%%%%%%%%%%%%%%%%%%%%%
%%%%%%%%%%%%%%%%%%%%%%%%%%%%%%%%%%%%%
\section{Tests of the No Evolution Scenario}
\label{sec:scaling_noev}

\tabletypesize{\scriptsize}
\begin{deluxetable*}{cccccc}
\tablewidth{0pt}
\tablecaption{Fit Parameters Assuming No Evolution}
\tablehead{
\multicolumn{6}{c}{Core Subtracted Relations} \cr
& & & & Diff. from 0 &  \\
Relation & $\alpha$ & $A$\tablenotemark{a} & $\gamma$ & (\%)\tablenotemark{b} & \sint\tablenotemark{c}
}
\startdata
%$^{+}_{-}$
\LXCStfive--\Txcs & $1.84^{+0.18}_{-0.14}$ & $1.74^{+0.26}_{-0.22}$E44 & $0.56^{+0.37}_{-0.35}$ & $88+$ & $0.24^{+0.04}_{-0.05}$ \\ [0.3em]
%  $0.11\pm0.02$
\Micmtfive--\Txcs  & $1.57\pm0.07$ & $2.40^{+0.17}_{-0.16}$E13 & $0.20^{+0.20}_{-0.12}$ & $89+$ & $0.00\pm0.00$   \\ [0.3em]
% $0.00\pm0.00$
\Micmfive--\Txcs & $1.52\pm0.08$ & $7.59^{+0.54}_{-0.51}$E13 & $0.10^{+0.19}_{-0.17}$ & $41+$ & $0.08\pm0.04$  \\ [0.3em]
% $0.04\pm0.02$

\tableline 
\multicolumn{6}{c}{Non-Core Subtracted Relations} \cr
& & & & Diff. from 0 & \\
Relation & $\alpha$ & $A$\tablenotemark{a} & $\gamma$ & (\%)\tablenotemark{b} & \sint\tablenotemark{c}   \\
\tableline \tableline 
%$^{+}_{-}$
\LXtfive--\Tx & $2.75^{+0.34}_{-0.25}$ & $2.81^{+0.73}_{-0.58}$E44 & $-0.25\pm0.56$  & $36-$ & $0.59\pm0.09$  \\ [0.3em]
%$0.26\pm0.04$
\Micmtfive--\Tx & $1.78\pm0.08$ & $2.57^{+0.18}_{-0.17}$E13 & $-0.10^{+0.17}_{-0.15}$ & $43-$ & $0.13\pm0.02$   \\ [0.3em]
% $0.06\pm0.01$
\Micmfive--\Tx & $1.74\pm0.10$ & $8.13^{+0.58}_{-0.54}$E13 & $-0.25^{+0.19}_{-0.18}$ & $78-$ & $0.14\pm0.02$ 
% $0.06\pm0.01$
\enddata
\tablenotetext{a}{In units of $L_\odot$ for \LX--\Tx\ relations, $M_\odot$ for \Micm--\Tx\ relations.}
\tablenotetext{b}{Significance level at which $\gamma$ differs from zero, as determined by bootstrap sampling and refitting; the sign indicates whether $\gamma$ is positive ($+$) or negative ($-$).}
\tablenotetext{c}{Intrinsic scatter in \LX\ or \Micm\ at fixed temperature, expressed in base $e$.}
\label{tab:noEfits}
\end{deluxetable*}

We now examine the evolution of scaling relations while assuming no evolution, i.e., we assume no scaling in \rd\ when measuring \LX\ and \Micm, and when determining the core subtraction radius for \Txcs\ and \LXCS. Our values for \LX\ and \Micm\ measured using the non-core subtracted temperature are given in Table~\ref{tab:noEinfo}. We do not measure \LXfive\ or \LXCSfive\ in this scenario, as only a handful of clusters have observations of sufficient exposure time and angular extent that we can measure out to the non-evolved \rfive.

 \begin{inlinefigure}
   {\myputfigure{lumpanel_noev}{2}{1.5}{-170}{-30}}
   \figcaption{\label{fig:lumpanel_noev} 
Projected X-ray luminosity within \rtfive\ (left) and \rfive (right), with non-core subtracted (top) and core subtracted (bottom) quantities, plotted versus temperature. These quantities are measured assuming {\it no evolution}. Luminosity values are scaled to $z=0$ using the best-fit redshift scaling for each relation, and the best-fit slope is plotted for each relation. Markers correspond to $z<0.4$ (circles), $0.4<z<0.6$ (squares), and $z>0.6$ (triangles).
}
\end{inlinefigure}

 \begin{figure*}
   {\myputfigure{masspanel_noev}{7}{.7}{-160}{-40}}
   \figcaption{\label{fig:masspanel_noev} 
Same as Figure~\ref{fig:lumpanel_noev}, but for \Micm--\Tx\ relations.
}
\end{figure*}

 \begin{figure*}[t]
   {\myputfigure{all_evol_noev_ydir}{7}{.9}{-330}{-60}}
   \figcaption{\label{fig:evol_noE} 
Ratio of measured observable (luminosity or ICM mass)  to the best-fit observable--temperature scaling relation, plotted versus redshift. These measurements assume {\it no evolution}. The horizontal line ($\mathcal{O}/\mathcal{O_{\rm fit}}=1$)  corresponds to no evolution, i.e., $\gamma=0$ in our notation. The dashed and dotted lines correspond to the best fit and $1 \sigma$ boundaries on $\gamma$ for each relation.
}
\end{figure*}

\subsection{Scaling Relations and Their Evolution}

The \LX--\Tx\ and \LX--\Micm\ relations are plotted in Figures~\ref{fig:lumpanel_noev} and \ref{fig:masspanel_noev}, respectively. As in the self-similar evolution case, the slope of the luminosity--temperature relation decreases significantly when core-subtracted quantities are used, and the scatter likewise decreases for both the luminosity and the ICM mass relations. We give the best-fit scaling relation parameters from this scenario in Table~\ref{tab:noEfits}, and plot the redshift evolution of the scaling relations in Figure~\ref{fig:evol_noE}.

The measured slopes and normalizations in this scenario are consistent with those measured in \S~\ref{sec:scaling_SS}, including the tendency for core subtracted relations to have shallower slopes than non-core subtracted relations. Also in common between the two scenarios is the tendency for core subtracted relations to have more positive evolution than non-core subtracted relations.

Single non-core subtracted relations are generally consistent with negative evolution, and core subtracted relations are generally consistent with positive evolution. Combining all three core subtracted relations gives a combined consistency with $\gamma=0$ (i.e., with the predictions of the no evolution scenario) of $1\%$; for the non-core subtracted relations, the value is 8\%.

\subsection{Summary of No Evolution Scenario Results}

The core subtracted scaling relations rule out the ``no evolution" scenario at 99\% confidence; non-core subtracted relations give less certain results. As in the self-similar evolution scenario, the core subtracted relations have slopes that are consistent with self-similar expectations, and evolution that is more rapid than the corresponding non-core subtracted relations. Together with the results from the self-similar evolution tests, these findings indicate that cluster scaling relations do evolve, but they evolve less rapidly than the self-similar expectation.

 \begin{figure*}
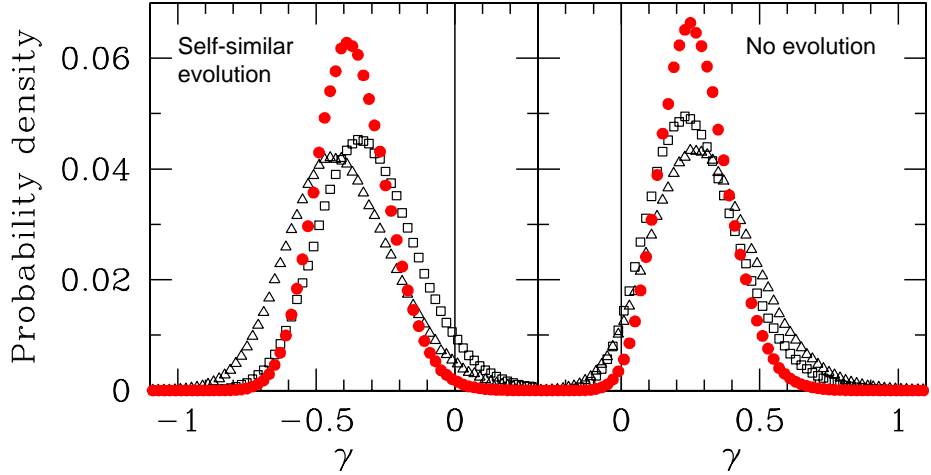

   {\myputfigure{fICMev_comb}{7}{.8}{-280}{-60}}
   \figcaption{\label{fig:fICMev} 
Constraints on the evolution of \ficm\ for the self-similar evolution ({\it left}) and no evolution ({\it right}) scenarios. Open triangles are from the fit to the \LXCStfive-\Txcs\ relation (with the values halved, as discussed in the text), open squares are from \Micmtfive-\Txcs\ relation, and filled circles are the normalized product of the two. The best-fit to the combined relations gives $\gamma_{f_g} = -0.39 \pm 0.13$ in the self-similar evolution scenario, and $\gamma_{f_g} = 0.25^{+0.12}_{-0.11}$ in the no evolution scenario.
}
\end{figure*}

 \begin{figure*}[t]
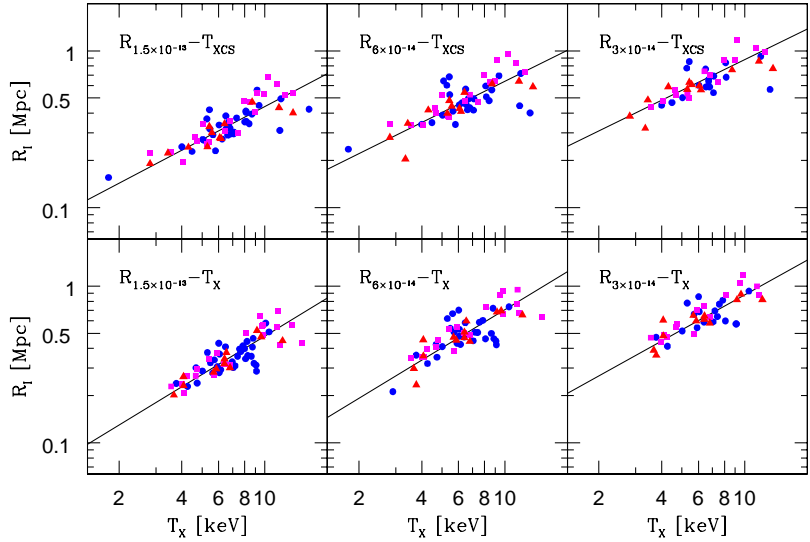

   {\myputfigure{sizepanel}{7}{.7}{-120}{-120}}
   \figcaption{\label{fig:sizepanel} 
Isophotal size--temperature relations for non-core subtracted (top) and core subtracted (bottom) temperature; the isophote used decreases from left to right. Size values are scaled to $z=0$ using the best-fit redshift scaling for each relation, and the best-fit slope is plotted for each relation. Markers vary by redshift as in Figure~\ref{fig:lumpanel}. 
}
\end{figure*}

 \begin{figure*}[t]
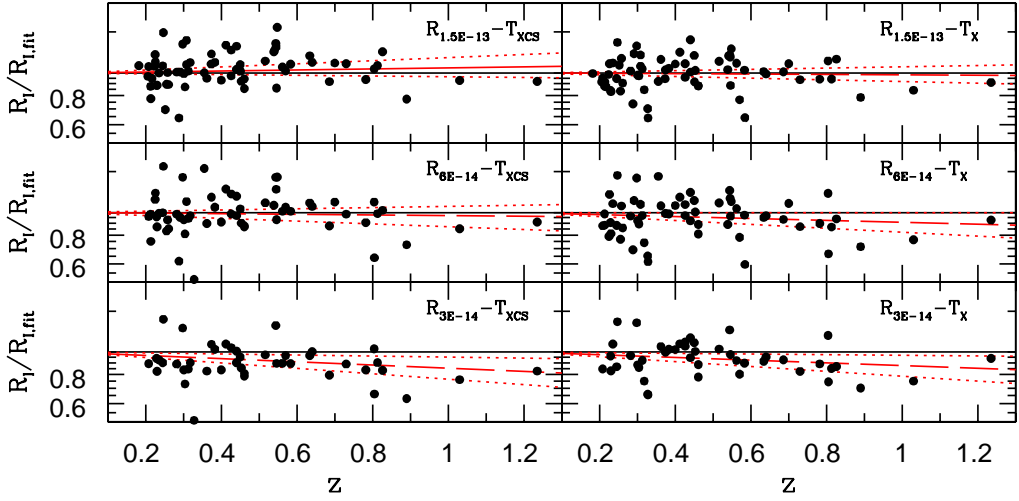

   {\myputfigure{RI_evol_ydir}{7}{.9}{-330}{-60}}
   \figcaption{\label{fig:evol_RI} 
Ratio of measured isophotal size to the best-fit  size--temperature scaling relation, plotted versus redshift. These measurements assume no evolution. The horizontal line ($R_I/R_{I, {\rm fit}}=1$)  corresponds to no evolution, i.e., $\gamma=0$ in our notation. The dashed and dotted lines correspond to the best fit and $1 \sigma$ boundaries on $\gamma$ for each relation.
}
\end{figure*}

%%%%%%%%%%%%%%%%%%%%%%%%%%%%%%%%%%%%%
%%%%%%%%%%%%%%%%%%%%%%%%%%%%%%%%%%%%%
\section{Testing Evolution of the ICM Fraction}
\label{sec:ficm}

One simple model for the evolution of cluster parameters such as \LX\ and \Micm\ is a simple evolution of the gas mass fraction \ficm, i.e., the ratio of the ICM mass to the total mass (baryons + dark matter) of a cluster. It is sometimes assumed in cosmological studies using clusters that \ficm\ is constant with redshift if clusters are selected appropriately \citep[e.g.,][]{rines99,allen04}, but this assumption is difficult to test because of degeneracies between \ficm\ measurements and cosmological parameters. Simulations disagree on the baryon fraction evolution, with some claiming to see a negative evolution \citep[e.g.,][]{kay07}, while others find no evolution \citep[e.g.,][]{crain07}.

We can test whether our data are consistent with an evolution in \ficm\ by directly combining measured values of $\gamma$ for individual scaling relations.  X-ray luminosity varies proportional to the square of the ICM density, and ICM mass is directly proportional to the ICM density. Because we are working in log space, this means that we combine $\gamma_{M_{g}}$ with $\gamma_{L_{\rm X}}/2$. 
We use the core subtracted relations for this test because these relations are presumably less biased by cluster structural changes in the core, and therefore more sensitive to more global changes in the gas fraction.

First we examine the \LXCStfive-- and \Micmtfive--\Txcs\ relations measured in the self-similar evolution scenario. The left panel of Figure~\ref{fig:fICMev} shows histograms for the values of $\gamma$ resulting from the bootstrap fitting of the \LXCStfive-- and \Micmtfive--\Tx\ relations (triangles and squares, respectively; the values of $\gamma$ for luminosity have been divided by 2 as explained above); the vertical axis has been scaled so that the values represent the probability of $\gamma$ falling in each bin. The circles are the product of the two individual distributions, renormalized so that the total probability is unity. The data give a best fit value of $\gamma_{f_g} = -0.39 \pm 0.13$; the data are inconsistent with $\gamma_{f_g}=0$ (i.e., a constant gas fraction) at the 99.1\% level. 

The right panel of Figure~\ref{fig:fICMev} shows data calculated in the same way, but in the no evolution scenario. In this scenario we find the best-fit combined scaling to be $\gamma_{f_g} = 0.25^{+0.12}_{-0.11}$, and inconsistent with $\gamma_{f_g}=0$ at the 98\% level.

Our results are consistent with the evolution in \LX\ and \Micm\ originating from a simple evolution in gas mass fraction. While such consistency does not prove  this scenario, it is encouraging to note that the values of $\gamma_{M_{g}}$ and $\gamma_{L_{\rm X}}/2$ are quite similar in both scenarios, and evolution in \ficm\ thus provides a consistent explanation for the evolution of these two different physical quantities. The most probable value $\gamma_{f_g} \simeq -0.4$ in the self-similar evolution scenario suggests a  decrease of $\sim$25\% in \ficm\ between redshifts 0 and 1, which would bias distance measurements that assume constant \ficm\ at the $\sim$17\% level \citep[$d_A\propto f_{g}^{2/3}$; e.g.,][]{rines99}.

Note that we have measured the evolution of \ficm\ specifically within the radius \rtfive. We do not attempt a similar measurement at \rfive\ because of a lack of luminosity measurements at that radius in the no evolution scenario, and the very large uncertainties on the \LXCSfive--\Txcs\ relation in the self-similar evolution scenario. There is in both scenarios and in both core subtracted and non-core subtracted relations a tendency for \Micmfive\ to evolve more slowly than \Micmtfive (though only at the 0.5--1$\sigma$ level); this is consistent with observations and simulations which find that the evolution in \ficm\ decreases with increasing radius, with evolution nearing zero at the virial radius \citep[e.g.,][]{sadat05,ettori06}.  This suggests that the distance biases associated with the assumption of constant gas fraction would be less severe if the X-ray data were deep enough to allow the measurements to be made at or beyond the virial radius, which may be possible which future observatories such as {\it Constellation-X}.

\tabletypesize{\scriptsize}
\begin{deluxetable*}{cccccc}
\tablewidth{0pt}
\tablecaption{Fit Parameters For Isophotal Size Relations}
\tablehead{
\multicolumn{6}{c}{Core Subtracted Relations} \cr
& & & & Diff. from 0 & \\
Relation & $\alpha$ & $A$ (Mpc) & $\gamma$ & (\%)\tablenotemark{a} & \sint\tablenotemark{b}
}
\startdata
%$^{+}_{-}$
\Rone--\Txcs & $0.70^{+0.07}_{-0.06}$ & $0.31\pm0.02$ & $0.08^{+0.17}_{-0.14}$ & $42+$ & $0.13\pm0.02$   \\ [0.3em]
% $0.06\pm0.01$
\Rsix--\Txcs &  $0.66^{+0.09}_{-0.08}$ & $0.46 \pm 0.03$ & $-0.05^{+0.15}_{-0.18}$ & $33-$ & $0.16\pm0.02$ \\ [0.3em]
% $0.07\pm0.01$
\Rthree--\Txcs & $0.65\pm0.10$ &  $0.63^{+0.05}_{-0.04}$ & $-0.26\pm0.18$ & $87-$ & $0.12^{+0.03}_{-0.02}$   \\ [0.3em]
% $0.05\pm0.01$

\tableline 
\multicolumn{6}{c}{Non-Core Subtracted Relations} \cr
& & & & Diff. from 0 &\\
Relation & $\alpha$ & $A$ (Mpc) & $\gamma$ & (\%)\tablenotemark{a} & \sint\tablenotemark{b} \\
\tableline \tableline 
%$^{+}_{-}$
\Rone--\Tx & $0.81\pm0.07$ & $0.32 \pm 0.01$ & $-0.03^{+0.13}_{-0.11}$ & $13-$ & $0.13 \pm 0.01$ \\ [0.3em]
% $0.06 \pm 0.01$
\Rsix--\Tx & $0.81 \pm 0.09$ & $0.47 \pm 0.03$ &  $-0.16 \pm 0.16$ & $68-$ & $0.16\pm0.02$  \\ [0.3em]
% $0.07\pm0.01$
\Rthree--\Tx & $0.74^{+0.08}_{-0.07}$ & $0.60^{+0.06}_{-0.04}$ & $-0.22 \pm0.17$ & $79-$ & $0.14 \pm0.02$
% $0.06 \pm0.01$
\enddata
\tablenotetext{a}{Significance level at which $\gamma$ differs from zero, as determined by bootstrap sampling and refitting; the sign indicates whether $\gamma$ is positive ($+$) or negative ($-$).}
\tablenotetext{b}{Intrinsic scatter in \RI\ at fixed temperature, expressed in base $e$.}
\label{tab:RIfits}
\end{deluxetable*}

%%%%%%%%%%%%%%%%%%%%%%%%%%%%%%%%%%%%%
%%%%%%%%%%%%%%%%%%%%%%%%%%%%%%%%%%%%%
\section{Evolution of Isophotal Size}
\label{sec:size}

We now examine the evolution of isophotal size--temperature scaling relations. This is done separately from the previous ``self-similar evolution" and ``no evolution" because as discussed in \S~\ref{sec:theory}, for clusters that are described by a $\beta$ model with $\beta=\frac{2}{3}$, the two scenarios give the same result \citep{mohr00}. That is, if the ICM is distributed as $r^{-2}$ beyond the core, the size--temperature relation does not evolve with redshift. While this potentially makes the size--temperature relation useful as a means to study the evolution of the angular diameter distance, and hence as a tool for studying cosmology, it makes it less useful for constraining the evolution of the ICM and cluster structure as we have done with luminosity and ICM mass relations.

\subsection{Scaling Relations and Their Evolution}

Size--temperature scaling relations are shown in Figure~\ref{fig:sizepanel}; as with the previous scaling relation plots, these have had the measured redshift evolution projected out. Best-fit scaling relation parameters are given in Table~\ref{tab:RIfits}. The slopes of the relations using core subtracted temperatures are consistent with the theoretical value $\alpha=\frac{2}{3}$ \citep{mohr00}, and the relations with non-core subtracted temperature are somewhat higher. Our  fit slope for the  \Rthree-\Tx\ relation is $0.74^{+0.08}_{-0.07}$, which differs significantly from the value $\alpha=0.93\pm0.11$ found by \citet{mohr00}  using \ROSAT\ PSPC images and literature values for \Tx.   

Redshift evolution of the isophotal size relations is shown in Figure~\ref{fig:evol_RI}. For the  fits to the entire sample, the isophotal size relations show little or no evolution in the isophote closest to the core, and a trend toward more negative evolution as the isophote used decreases, i.e., as one examines the cluster at distances further from the core. 

Having shown in \S~\ref{sec:ficm} that the evolution in \LX\ and \Micm\ with respect to the self-similar expectation can be modeled by a simple evolution in the gas fraction, we can check for consistency of that evolution with the isophotal size results. The brightness at a given cluster radius $r$ is related to the gas fraction \ficm\ as $I(r) \propto f_{g}^2$, and so it can be shown that for a cluster described by a spherical $\beta$ model the measured isophotal size scales with $I(r)$ as $R_I \propto I(r)^{1/(6\beta-1)}$ \citep{mohr00}. Thus we expect

\begin{equation}
R_I \propto f_{g}^{2/(6\beta-1)},
\end{equation}

\noindent
which, for the standard value of $\beta=\frac{2}{3}$  \citep[e.g.,][]{jones84,mohr99}, means that isophotal size should scale as $f_g^{2/3}$. For our self-similar evolution measurement of $\gamma_{f_g} = -0.39 \pm 0.13$, this would predict $R_I \propto (1+z)^{-0.26\pm0.09}$, in good agreement with the directly measured evolution  of $\gamma = -0.26\pm0.18$ in the \Rthree--\Txcs\ relation, and  of $\gamma = -0.05^{+0.15}_{-0.18}$ in the \Rsix--\Txcs\ relation. More positive evolution at higher isophotes may be an indication of structural changes as clusters evolve and the density profiles of clusters become more peaked at lower redshift.

 \begin{inlinefigure}
   {\myputfigure{plot_dA_resid}{.8}{1.3}{-135}{-50}}
   \figcaption{\label{fig:plot_dA} 
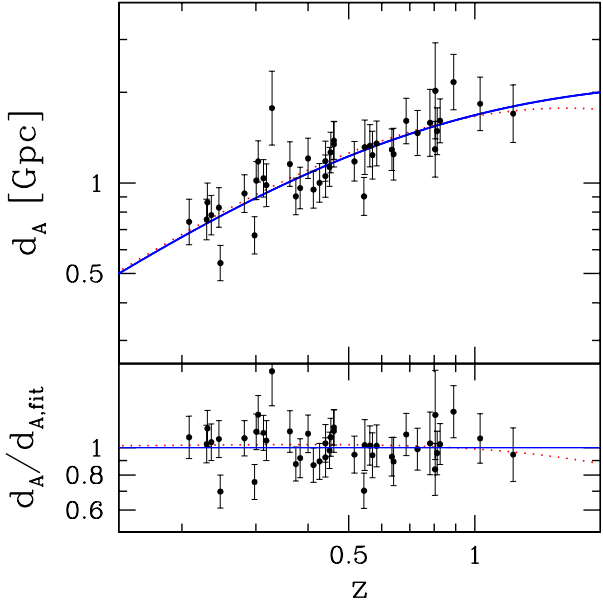
Measured angular diameter distance $d_A$ plotted versus redshift for the \Rthree\ sample. The solid line shows $d_A(z)$ for  the best-fit measured cosmology $\Omega_M = 0.02$, $\Omega_\Lambda = 0.31$, and the dotted line shows $d_A(z)$ for our adopted cosmology $\Omega_M = 0.266, ~\Omega_\Lambda = 0.734$.
}
\end{inlinefigure}

\subsection{Prospects for Cosmology Using Isophotal Size}
\label{sec:sizecosmo}

As mentioned above, the predicted non-evolution of \RI\ with redshift makes these size measurements a promising source of angular diameter distances, which can be used to constrain cosmological parameters. Such an undertaking is beyond the scope of this paper, but we sketch here the basic ideas underlying such a measurement.

If isophotal size indeed evolves in a manner predictable by the evolution in \ficm, then one can use a measured angular isophotal size $\theta_I$, together with a physical isophotal size $R_I$ for the same cluster predicted from a scaling relation, to determine the angular diameter distance, $d_A = R_I/\theta_I$; this can then be used to measure the cosmological parameters which determine $d_A(z)$. As a test, we use $\theta_I$ measured from our \Rthree\ sample, and use the best-fit slope and normalization found for the \Rthree-\Txcs\ relation to predict $R_I(T_{\rm X},z)$. Because we have found evolution in \LX\ and \Micm\ which suggests evolution in \ficm, we adopt  the best-fit \ficm\ evolution $\gamma_{f_g} = -0.39$ and its consequent isophotal size evolution $\gamma_{R_I} = -0.26$ in the size--temperature relation, as discussed above. Uncertainties in $d_A$ are a combination of the temperature uncertainty and the measured intrinsic scatter in the \Rthree-\Txcs\ relation. 

Note that this is not simply an independent cosmological test. This is a consistency test where (1) an input cosmological model is assumed, (2) \ficm\ evolution is measured using the evolution-sensitive \LX\ and \Micm\ scaling relations, and (3) that evolution is adopted in using isophotal sizes to derive an output cosmology. Because the cosmological dependencies of each scaling relation differ, the input and output cosmologies will only agree for the correct model.

Figure~\ref{fig:plot_dA} shows the angular diameter distance versus redshift, with our input cosmology and the output best-fit cosmology. As can be seen, these data do not reach redshifts high enough to place tight constraints on cosmology. Figure~\ref{fig:omegaplot} shows confidence intervals for the density parameters \OmegaM\ and \OmegaL\ (we fix $H_0$ to our assumed value of 70.9 km~s$^{-1}$~Mpc$^{-1}$). The uncertainties on both parameters are quite large; fully marginalized constraints are $\Omega_M = 0.02^{+0.49}_{-0.02}$, $\Omega_\Lambda = 0.31^{+0.59}_{-0.19}$. We do successfully recover our input cosmology within the 1 $\sigma$ region.

This combination of the use of \LX--\Tx\ and \Micm--\Tx\ relations to constrain the evolution of the ICM, and \RI--\Tx\ relations to measure distances is an approach that deserves further attention. As X-ray surveys that include spectroscopic temperature measurements push to higher redshifts, the use of isophotal sizes to measure angular diameter distances as demonstrated here should provide a new source of cosmological measurements, complementary to other cluster methods and to CMB and supernova constraints.

 \begin{inlinefigure}
   {\myputfigure{OmegaPlot}{2.2}{1.8}{-290}{-70}}
   \figcaption{\label{fig:omegaplot} 
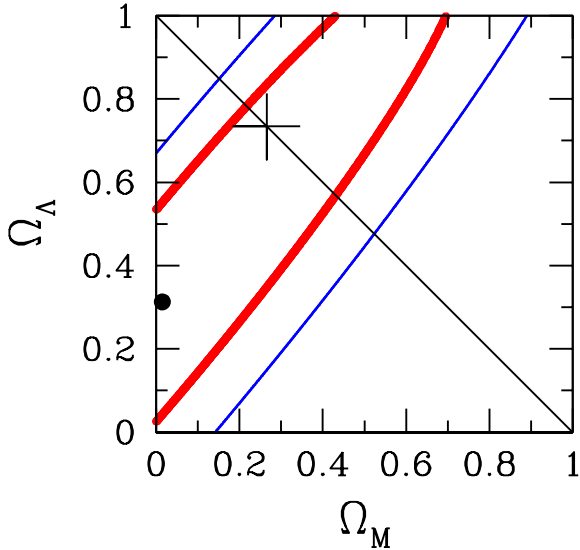
Constraints on \OmegaM\ and \OmegaL\ from fitting the angular diameter distances determined from isophotal sizes. The thick and thin contours mark the boundaries of the 1 and 2$\sigma$ confidence regions, respectively. The circle denotes the best fit $\Omega_M = 0.02$, $\Omega_\Lambda = 0.31$, and the cross marks our adopted cosmology for this paper, $\Omega_M = 0.266, ~\Omega_\Lambda = 0.734$.
}
\end{inlinefigure}

\tabletypesize{\scriptsize}
\begin{deluxetable*}{cccccccccc}
\tablewidth{0pt}
\tablecaption{Scatter and Slope Comparisons, Core Subtracted vs. 3-Parameter}
\tablehead{
& \multicolumn{2}{c}{Standard Relation} & & \multicolumn{2}{c}{Core Subtracted} & & \multicolumn{3}{c}{3-Parameter} \cr
\cline{2-3} \cline{5-6} \cline{8-10}
Relation & $\alpha$ & \sint & & $\alpha$ & \sint & & $\alpha$ & $\beta$ & \sint
}
\startdata
\LXtfive--\Tx & $2.75^{+0.29}_{-0.26}$ & $0.60^{+0.08}_{-0.09}$ && $2.00^{+0.23}_{-0.19}$ &  $0.28\pm0.05$  && $1.72\pm0.13$ & $0.39\pm0.03$ &$0.21\pm0.03$ \\ [0.3em]
\LXfive--\Tx & $2.35^{+0.33}_{-0.24}$ & $0.39^{+0.12}_{-0.10}$ && $2.26^{+0.29}_{-0.33}$ & $0.21^{+0.08}_{-0.07}$ && $1.72\pm0.13$ & $0.28^{+0.06}_{-0.04}$ &$0.15^{+0.02}_{-0.07}$ \\ [0.3em]
\Micmtfive--\Tx  & $1.82\pm0.08$ & $0.14\pm0.02$ && $1.63^{+0.09}_{-0.08}$ & $0.00^{+0.05}_{-0.00}$ && $1.70\pm0.07$ & $0.10 \pm 0.02$ & $0.07\pm0.02$ \\ [0.3em]
\Micmfive--\Tx & $1.74\pm0.09$ & $0.13\pm0.02$ && $1.56\pm0.10$ & $0.09 \pm 0.04$ && $1.70^{+0.10}_{-0.07}$ & $0.02\pm0.02$ &$0.12\pm0.02$ \\ [0.3em]
\Rone--\Tx & $0.81\pm0.07$ & $0.13 \pm 0.01$ && $0.70^{+0.07}_{-0.06}$ & $0.13\pm0.02$ && $0.89 \pm 0.09$ & $0.01\pm0.02$ & $0.13\pm0.02$ \\ [0.3em]
\Rsix--\Tx &  $0.81 \pm 0.09$ &  $0.16\pm0.02$ && $0.66^{+0.09}_{-0.08}$ & $0.16\pm0.02$ && $0.82 \pm 0.09$ & $-0.02^{+0.03}_{-0.02 }$ &$0.16\pm0.02$ \\ [0.3em]
\Rthree--\Tx & $0.74^{+0.08}_{-0.07}$ & $0.14 \pm0.02$ && $0.65\pm0.10$ & $0.12^{+0.03}_{-0.02}$ && $0.76\pm0.08$ & $-0.03\pm0.02$ & $0.13\pm0.02$
\enddata
\tablecomments{Scatter is given in base $e$.}
\label{tab:scattercomp}
\end{deluxetable*}

%%%%%%%%%%%%%%%%%%%%%%%%%%%%%%%%%%%%%
%%%%%%%%%%%%%%%%%%%%%%%%%%%%%%%%%%%%%
\section{Scatter in Scaling Relations}
\label{sec:scatter}

This paper has focused on the evolution of the normalization of observable--temperature scaling relations. Here we briefly discuss the {\it scatter} about those scaling relations, i.e., the variation in the ICM distribution from cluster to cluster at fixed temperature.  Understanding the precise origins of scatter helps both in understanding cluster physics such as cool core development and merger effects, and in understanding sources of uncertainty in cosmological studies that use observables such as X-ray luminosity and temperature as proxies for cluster mass. As shown by \citet{ohara06}, the cluster central surface brightness \Io\ is strongly correlated with central cooling time and reflects the core structure of clusters. In this section we examine the use of \Io\ to reduce scatter in scaling relations, and to examine the redshift evolution of cluster structure.

\subsection{Reducing Scatter: Two Approaches}
\label{sec:reduce}

As shown in previous sections, the total scatter in scaling relations generally decreases when core-subtracted quantities are used, reflecting the separation in cool core and non-cool core populations that is observed in most scaling relations \citep[e.g.,][]{fabian94c,markevitch98b,mccarthy04,ohara06}. \citet{ohara06} demonstrated that central surface brightness \Io\ can be used as a proxy for cool core ``strength" in a three parameter (\OTI) scaling relation, reducing the scatter in scaling relations that is introduced by  biases to both the temperature and to the other observable (\LX, \Micm, \RI) in the relation. With the data presented here we can compare the three-parameter approach to the use of core subtracted quantities, to determine whether either method results in lower scatter than the other.

Rather than using the $\beta$ model values for \Io, as in \citet{ohara06}, we estimate \Io\ by simply averaging the surface brightness within 0.05\rfive\ of the brightness peak. Since our intention is to use \Io\ to parametrize the development of cool cores, this method is likely to give more accurate results than the surface brightness fitting which, even when a double $\beta$ model is used, may not accurately reflect the structure around the brightness peak of a non-spherically symmetric cluster. We fit a scaling relation of the form

\begin{equation}
\mathcal{O} \propto T_{\rm X}^\alpha I_0^\beta(1+z)^\gamma,
\end{equation}

\noindent
using the orthogonal fit (Eq.~\ref{eq:orthfit}) appropriately modified for the additional parameter.

Table~\ref{tab:scattercomp} gives the \Tx\ dependence and intrinsic scatter for seven relations using non-core subtracted quantities (e.g., \LX--\Tx), for the same relations using core subtracted quantities (e.g., \LXCS--\Txcs), and for the same relations adding the third parameter \Io\ (e.g., \LX--\Tx--\Io). The 3-parameter \LX\ relations have even lower intrinsic scatter than the core subtracted relations; for the \Micm\ relations, the reverse is true. The scatter is little different between the different methods for the isophotal size relations, with perhaps slightly lower scatter in the core subtracted relations.

Interestingly, the slopes for the 3-parameter \LX\ relations are even lower than those of the core subtracted relations, and are $\sim2\sigma$ lower than the self-similar expectation $\alpha=2$. For the \Micm\ and \RI\ relations, however, the 3-parameter slopes are consistent with those of the original relation, i.e., steeper than the core subtracted relations.

The \Micm\ and \RI\ results by themselves would suggest that the three-parameter fit does not remove cool core-induced average temperature biases as completely as using core subtracted temperatures does; i.e., the brightness of a cluster's core is not a perfect indicator of the strength of the core. The reduced scatter in the three-parameter \LX\ relations compared with the core subtracted relations, however, indicates that differences in cool core and non-cool core clusters persist outside the 0.2\rfive\ core exclusion radius.
Together, these results may lend some additional weight to the argument that cool core and non-cool core clusters differ in ways other than their apparent relaxation as determined by the development of a cool, dense core.

 \begin{inlinefigure}
   {\myputfigure{I0panel}{2}{1.7}{-190}{-170}}
   \figcaption{\label{fig:I0panel} 
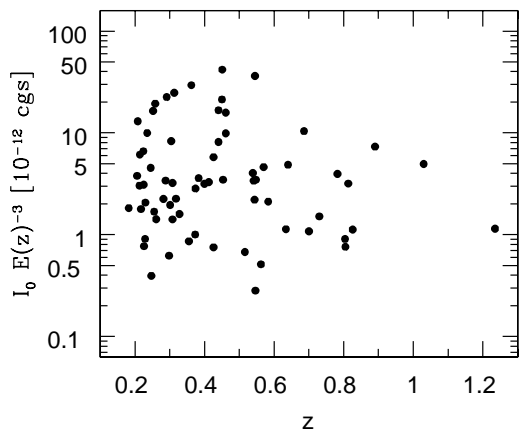
Central surface brightness \Io\  versus redshift. The values of \Io\ have been scaled  by $E(z)^{-3}$, as described in the text, and are given in cgs units, i.e., erg s$^{-1}$ cm$^{-2}$ arcmin$^{-2}$.
}
\end{inlinefigure}

\subsection{Evolution of Scatter}

As mentioned in \S~\ref{sec:Eevol}, we see a qualitative decrease in scatter at higher redshifts. \citet{kay07} found a decrease of a factor of $\sim$3 in the luminosity--temperature relation in simulations, which they ascribe to merger effects at lower redshift. However, observational studies have found that clusters are in fact more structurally disturbed at higher redshift \citep[e.g.,][]{jeltema05}. 
Further more, cool cores are nearly ubiquitous in the \citet{kay07} simulations at all redshifts, in contrast to observational results that find a fairly constant cool core fraction of $\sim$50\% up to $z=0.4$ \citep{bauer05}; \citet{ohara06} showed that cool core-related effects, and not mergers, are the primary contributors to scaling relation scatter at low redshift, and so clearly accurate simulation of core evolution is required if simulations are to constrain the evolution of this scatter. 

One way of gauging the effects of cool core development on scaling relation scatter is to look at the evolution of the central surface brightness \Io. In Figure~\ref{fig:I0panel} we plot \Io, measured as described in \S~\ref{sec:reduce},  redshift.
Like other cluster observables, \Io\ should evolve with redshift as clusters grow and the average density drops with the cosmic expansion. Because \Io\ is a measurement of the emission from a cluster along the line of sight through its center, i.e.,

\begin{equation}
I_0 \propto \int n_e^2 ~dr,
\end{equation}

\noindent
and density depends on redshift as as \Ez$^2$, and cluster radius depends on redshift as \Ez$^{-1}$, we expect $I_0 \propto E(z)^3$ if clusters evolve self-similarly. Thus the values of \Io\ in Figure~\ref{fig:I0panel} are scaled by $E(z)^{-3}$, and if clusters evolve self-similarly we would expect no average change with redshift in $I_0 E(z)^{-3}$ as plotted.

Qualitatively, however, it appears that the clusters with the highest \Io\ appear at low redshift, indicating a change in core structure at these redshifts. This is consistent with our findings that scaling relations with core subtracted quantities evolve faster with redshift than those with non-core subtracted quantities.  Furthermore, the overall scatter appears to increase at lower redshifts, consistent with what we have found in observable--temperature relations, indicating a wider range of core and other structural variations as clusters develop. Together, these trends can be explained by an increasing cool core fraction, or an increase the the strengths of cool cores in those clusters that have them, at lower redshifts. A larger sample of clusters would enable a more definitive investigation.

%%%%%%%%%%%%%%%%%%%%%%%%%%%%%%%%%%%%%
%%%%%%%%%%%%%%%%%%%%%%%%%%%%%%%%%%%%%
\section{Discussion}
\label{sec:discuss}

Our study indicates that cluster evolution is inconsistent with the simple self-similar model of cluster formation via gravitational collapse with no other heating or cooling processes. There is a substantial body of observational work in this area already, so in this section we discuss the similarities and differences between our work and earlier studies of scaling relation evolution. The ultimate goal of such observations is to constrain models of cluster formation; predictions of how cluster evolution will be modified by non-gravitational processes can be made both via simple analytical models \citep[e.g.,][]{voit05b} and from detailed hydrodynamical simulations \citep[e.g.,][]{muanwong06}.

\subsection{Luminosity--Temperature}

The X-ray luminosity--temperature relation is by far the most studied cluster scaling relation, with several studies using \Chandra\ or XMM. 
 These studies have generally found evolution in  \LX--\Tx\ relations that is either consistent with the self-similar expectation \citep[e.g.,][]{vikhlinin02,lumb04,kotov05,maughan06} or more negative \citep[e.g.,][]{ettori04a,branchesi07b}. An interesting exception is \citet{morandi07}, who found positive evolution when using their entire 24 cluster sample, but marginally negative evolution when using only the 11 clusters which were identified as having cool cores. 

Qualitative examination of the redshift scaling in our sample (Figures~\ref{fig:evol_E}, \ref{fig:evol_noE}, and \ref{fig:evol_RI}) clearly indicates the need to include clusters at redshifts as high as possible. Of the other studies mentioned above, the only ones that extend to redshifts beyond $z=0.8$ are \citet{ettori04a} and \citet{branchesi07b}, who find negative evolution with respect to self-similar, as we do; \citet{vikhlinin02}, who see no evolution with respect to self-similar, but whose methods (e.g., measurement of  luminosities within fixed 2 Mpc apertures) are quite different from later studies, making comparison difficult; and \citet{maughan06}, whose result is only marginally consistent with the self-similar expectation.

The work of \citet{branchesi07b} in particular is interesting to compare to ours, because they use a \Chandra\ sample covering a similar redshift range (though with only 17 members), and study two scenarios similar to our self-similar and no evolution scenarios. They find negative evolution with respect to self-similar, though at lower significance than our result; with an additional 22 clusters from three other \Chandra\ studies, the significance increases. In a no evolution scenario, they find the \LX--\Tx\ relation evolution to be consistent with zero, as we do in our \LXtfive--\Tx\ relation, which is most directly comparable. However, \citet{branchesi07b} additionally measure scaling with respect to slopes and normalizations from local relations, obtain poor fits, and conclude that there is different evolution in the luminosity--temperature relation between $0<z\lesssim0.3$ and above this range. As discussed, however, there are systematic differences between cluster parameters measured with different instruments, or even the same instrument in different studies, as is shown in the \citet{branchesi07b} results where fits worsen as additional clusters are added from other \Chandra\ studies. If there is a sharp change at low redshift, quantifying it will require a homogeneously reduced sample, a task made unfortunately difficult for \Chandra\ by its small field of view.

Results from simulations suggest possible explanations for the slower than self-similar evolution that we observe in the \LX--\Tx\ relation. While not trying to exactly duplicate observed relations, \citet{muanwong06} produced simulations using different models for the increase in entropy of the ICM. Their results show that, as naively expected, a simple radiative cooling model results in faster than self-similar evolution in luminosity--temperature because of reduced mean cluster temperatures and increased luminosities. They found slower than self-similar evolution using simple preheating and stellar feedback models, with the latter's negative evolution significantly greater than the former. While their models are simple and cannot be directly used to test specific realistic models, these results do illustrate the usefulness of scaling relations in constraining cluster physics.

\citet{ettori04b} and \citet{kay07} have studied scaling relation evolution in  simulations that include radiative cooling, star formation, and feedback. Both studies found significant ($\gg3\sigma$) negative evolution with respect to self-similar in bolometric \LXfive--\Tx\ relations; specifically, \citet{ettori04b} found $\gamma = -0.76 \pm 0.08$ (depending on the exact method used; the other possible values are the same within the uncertainty), and \citet{kay07} found $\gamma = -0.98 \pm0.03$ when using non-core subtracted quantities, and $\gamma = -0.61 \pm 0.04$ when measuring luminosities and temperatures excluding the central 50 kpc. Though direct comparisons may not be possible given differences in measurement of cluster temperatures between simulation and observation, differences in how the luminosities are measured, and the fact that our \LXfive\ samples are relatively small and consequently have large uncertainties in their fit parameters, the simulation results are consistent with our results in Table~\ref{tab:SSfits} for \LXfive\ and \LXtfive\ relations. The more negative scaling in the non-core subtracted relation that \citet{kay07} found in simulations is matched by our data, and indicates that the primary source of the slower than self-similar evolution in the \LX--\Tx\ relation is due to clusters being underluminous at higher redshifts, and not to temperature biases from cores. This slower than expected increase in luminosity at high redshifts indicates a potential source of difficulty for X-ray cosmology surveys, in that it may be more difficult to find large numbers of high-redshift clusters than has generally been assumed.

\subsection{ICM Mass--Temperature}

The ICM mass--temperature relation is less well studied than luminosity--temperature, and results are more varied. \citet{vikhlinin02} found significantly positive evolution relative to the self-similar expectation when measuring masses within a radius defined in terms of the average baryon density of the Universe; \citet{ettori04a} found marginally significant (1--2 $\sigma$) negative evolution with respect to self-similar ($\gamma =  -$(0.1--0.4), depending on the method used); \citet{maughan06} claim consistency of their high-redshift sample with low-redshift clusters when self-similar scaling is applied, though they do not attempt to directly measure any evolution; and \citet{morandi07} find significantly positive evolution with respect to self-similar.  The simulations of \citet{ettori04b} predict negative evolution ($\gamma=-$(0.1--0.2), depending on the method) at the 1--2 $\sigma$ level.   To this we compare our results, in which we find that \Micm\ has negative evolution with respect to self-similar at the 1--3 $\sigma$ level, depending on the radius  and whether  core subtracted parameters are used.

\subsection{Gas Fraction}

An unchanging gas mass fraction, or one that changes in easily quantifiable ways, is an essential component of cosmological studies that use measurements of cluster gas mass fractions to study cosmology \citep[e.g.,][]{rines99,ettori03,allen04,allen07}. There are, however, several complications to this use of \ficm, which varies by cluster mass and by radius within a cluster \citep[e.g.,][]{david95,mohr99,sanderson03,sadat05}. \citet{sadat05} claimed to find a decrease in \ficm\ at higher redshifts when assuming a standard $\Lambda$CDM cosmology, consistent with our findings that \ficm\ within \rtfive\ decreases with redshift relative to the self-similar expectation. The angular diameter distance of clusters, which is used in these cosmological studies, varies with \ficm\ as $d_A \propto f_{g}^{2/3}$, and so our observed $\sim$25\% decrease in \ficm\ between redshifts 0 and 1 corresponds to a bias of $\sim$17\% in $d_A$ estimates based on constant \ficm\ over the same redshift range.

 Simulations that include radiative cooling, star formation, and feedback processes likewise predict this decrease in \ficm\ with redshift, with the magnitude of that decrease being larger at smaller fractions of the cluster virial radius \citep[e.g.,][]{kravtsov05,ettori06}. As with the \LX\ and \Micm\ evolution, the predicted magnitude of this evolution differs according to the simulation parameters and the numerical codes used \citep{ettori06}, and so observational results such as ours will provide constraints as simulation quality improves. 

As has been demonstrated \citet[][]{ferramacho07}, the results obtained from cosmological studies that assume constant gas fraction depend heavily on the radius within which measurements are made, with radii closer to the virial radius giving results that disagree greatly with the concordance model. Though measurements at large radii require extrapolation that may introduce additional biases, such results when combined with evidence of the radial and redshift dependence of \ficm\ give strong warning against ready acceptance of cosmological results that assume constant \ficm, particularly when measurements are made at small radii such as \rtfive. 

Though our results suggest difficulties for cosmological studies that assume constant \ficm, we have presented in \S\ref{sec:sizecosmo} an alternative method for studying cosmology that involves using information about the evolution of \ficm\ to measure angular diameter distances using isophotal sizes. 
This cosmic consistency test requires joint analysis of cluster structure, using \LX\ and \Micm\ to constrain ICM evolution and \RI\ to estimate distances. Consistent input and output cosmological models are guaranteed only around the correct model.

%%%%%%%%%%%%%%%%%%%%%%%%%%%%%%%%%%%%%
%%%%%%%%%%%%%%%%%%%%%%%%%%%%%%%%%%%%%
\section{Conclusions}
\label{sec:concl}

We study the evolution of the ICM using X-ray scaling relations measured from a large, homogeneously analyzed sample of clusters spanning $0.2 \lesssim z \lesssim 1.2$. We use luminosity-- and ICM mass-temperature relations, including both relations with and without core subtracted quantities, to test scenarios of standard ``self-similar evolution" and of ``no evolution". We also study the evolution of isophotal size--temperature relations, for which (under certain assumptions) these two scenarios are identical. Finally, we compare the scatter in scaling relations after attempting to reduce cool core-induced scatter in two different ways. Our principal results appear below:

\begin{enumerate}

\item{Luminosity-- and ICM mass-temperature relations evolve less rapidly than expected in the self-similar evolution scenario; that is, clusters at higher redshifts have systematically lower luminosity and ICM mass at a given temperature than would be expected if clusters evolved self-similarly. The core subtracted relations have a combined consistency with the self-similar prediction of $<$0.1\%; non-core subtracted relations are even more inconsistent with the self-similar prediction.} 

\item{The data are also inconsistent with the no evolution scenario, though not at as strongly as in the self-similar scenario. The core subtracted relations  evolve more rapidly than expected at higher redshift in this scenario, with combined probability of consistency with no evolution of 1\%.}

\item{The evolution in the \LXCS--\Txcs\ and \Micm--\Txcs\ relations is consistent with a simple evolution in gas fraction, with  evolution in \ficm\  at $>99$\% confidence ($\gamma_{f_g} = -0.39 \pm 0.13$) in  the self-similar evolution   scenario when using core subtracted observables measured within \rtfive.}

\item{Isophotal size evolves with redshift at a rate that depends on the isophote used, reflecting evolution in the ICM spatial distribution in clusters. Evolution of isophotal size at a low isophote (i.e., well away from the core) is consistent with that expected given the measured \ficm\ evolution.}

\item{Relations with core subtracted quantities in general have more positive evolution than relations with the cores included, suggesting that either the cool core fraction decreases with increasing redshift, or that the cool core fraction remains constant but the cores that do exist are weaker at high redshift. This is supported by direct observations of the redshift dependence of central surface brightness, a good indicator of cool core development; the scatter and magnitude of \Io\  increase at low redshift.}

\item{Core subtracted relations generally have temperature dependences that are shallower than non-core subtracted relations, and thus are more consistent with the slopes predicted by the self-similar model for each scaling relation.}

\item{The use of core subtracted quantities for scaling relations and the use of non-core subtracted quantities with the addition of a third parameter, the central surface brightness, both significantly reduce scaling relation scatter by compensating to some extent for cool core-related effects. }

\item{Scatter in observables at fixed temperature appears to decrease with redshift. This could indicate an increase in the cool core fraction, an increase in the strength of cool cores in those clusters that have them, or both.}

\end{enumerate}

Cluster simulations are still improving with regard to their ability to accurately model non-gravitational processes and thus to directly test specific models by comparison to observational data. However, our results of negative evolution with respect to self-similar expectations in \LX\ and \Micm, and consequently in \ficm, provide important constraints for future computational studies. Our findings provide new warnings with regard to the assumptions made when using \ficm\ measurements to study cosmology. It has long   been established that \ficm\ varies with radius inside clusters and   varies with cluster mass when measured within \rfive\ \citep[e.g.,][]{david93,mohr99}.  Our results strongly suggest that \ficm\ varies with redshift as well.  Given the differences in behavior of collisionless   dark matter and the ICM (particularly the ICM's sensitivity to radiative   cooling and feedback from AGN and supernovae), perhaps it should not be   surprising that these components vary differentially with radius,   cluster mass, and even redshift.

At the same time, the combination of isophotal size measurements with measurements of the evolution of \ficm\ from \LX\ and \Micm\ relations provides a promising tool for measuring angular diameter distances. Our proposed cosmic consistency test would   allow one to use cluster structure and its evolution to constrain   cosmology in a manner complementary to more established techniques.   Finally, our results underscore the need to directly calibrate (or   self-calibrate) mass--observable scaling relations in large cluster   survey cosmology experiments.  Cluster structural evolution is   subject to a wide range of interesting physics, and determining that   mix reliably enough for even the most sophisticated simulations to   precisely predict cluster mass--observable scaling relations and   their evolution will remain enormously challenging for the   foreseeable future.

%%%%%%%%%%%%%%%%%%%%%%%%%%%%%%%%%%
%%%%%%%%%%%%%%%%%%%%%%%%%%%%%%%%%%
\acknowledgments

This work was supported by the NASA Long Term Space Astrophysics award NAG 5-11415, NASA GSRP fellowship NNG05GO42H, and NSF Grant No. 0611808.

\bibliographystyle{apj}
\bibliography{cosmology,evol}

\tabletypesize{\tiny}
\begin{deluxetable*}{p{2.2cm}ccccccccc}

\tablewidth{0pt}
\tabletypesize{\tiny}
\tablecaption{Observation and Spectral Fitting Information}

\tablehead{
Cluster & $z$ &
\colhead{ObsID} & $t_{\rm exp}$\tablenotemark{a} & \colhead{RA\tablenotemark{b}} & \colhead{DEC\tablenotemark{b}} & \colhead{$T_{\rm X}$ aperture} &
\colhead{\Tx} & \colhead{\Txcs, SS ev.\tablenotemark{c}} & \colhead{\Txcs, no ev.\tablenotemark{d}} \\ 
 & & & (ks) & &  & \colhead{(arcsec)} & \colhead{(keV)} & \colhead{(keV)} & 
 \colhead{(keV)}  
}
\startdata
A665\dotfill & 0.182 & 3586 & 29.6 & 08:30:50.2 & $+$65:52:14 & 380 &	8.0$\pm$0.2 &	8.1$\pm$0.3 &	8.2$\pm$0.3	\\
A963\dotfill & 0.206 & 903 & 29.9 & 10:17:03.8 & $+$39:02:42 & 195 &	7.0$\pm$0.3 &	6.8$^{+0.4}_{-0.5}$ &	6.8$^{+0.4}_{-0.5}$ 	\\
RX J0439.0+0520\dotfill & 0.208 & 527 & 9.6 & 04:39:02.3 & $+$05:20:45 & 204 &	4.3$^{+0.4}_{-0.3}$ &	4.0$^{+0.9}_{-0.6}$ &	4.0$^{+1.1}_{-0.6}$ 	\\
A1423\dotfill & 0.213 & 538 & 9.7 & 11:57:18.1 & $+$33:36:45 & 256 &	6.0$\pm$0.4 &	6.2$\pm$0.7 &	6.3$^{+0.8}_{-0.7}$ 	\\
ZwCl 2701\dotfill & 0.214 & 3195 & 18.3 & 09:52:49.3 & $+$51:53:05 & 150 &	4.7$\pm$0.2 &	5.8$\pm$0.6 &	6.0$^{+0.7}_{-0.6}$ 	\\
A773\dotfill & 0.217 & 5006 & 19.8 & 09:17:53.0 & $+$51:43:37 & 257 &	8.3$\pm$0.4 &	8.0$\pm$0.6 &	8.1$^{+0.7}_{-0.6}$ 	\\
A2261\dotfill & 0.224 & 5007 & 24.3 & 17:22:27.1 & $+$32:07:56 & 275 &	7.7$^{+0.3}_{-0.2}$ &	7.3$\pm$0.5 &	7.1$\pm$0.5	\\
ACO 2246\dotfill & 0.225 & 547 & 48.2 & 17:00:41.5 & $+$64:12:53 & 103 &	2.9$^{+0.3}_{-0.2}$ &	1.8$^{+0.3}_{-0.2}$ &	1.7$^{+0.3}_{-0.2}$ 	\\
A1682\dotfill & 0.226 & 3244 & 4.7 & 13:06:55.1 & $+$46:33:01 & 254 &	5.5$^{+0.8}_{-0.4}$ &	5.4$^{+1.0}_{-0.6}$ &	5.6$^{+1.1}_{-0.6}$ 	\\
A2111\dotfill & 0.229 & 544 & 10.2 & 15:39:39.6 & $+$34:25:55 & 298 &	7.2$\pm$0.7 &	7.1$^{+1.0}_{-0.9}$ &	6.6$^{+1.1}_{-0.6}$ 	\\
A267\dotfill & 0.230 & 3580 & 19.9 & 01:52:42.1 & $+$01:00:33 & 254 &	7.1$^{+0.4}_{-0.5}$ &	6.8$^{+1.1}_{-0.5}$ &	7.1$^{+0.9}_{-0.8}$ 	\\
RX J2129.7+0005\dotfill & 0.235 & 552 & 9.9 & 21:29:40.1 & $+$00:05:18 & 218 &	5.7$\pm$0.3 &	6.7$^{+1.1}_{-0.6}$ &	6.8$^{+1.3}_{-0.6}$ 	\\
RX J0439.0+0715\dotfill & 0.245 & 3583 & 19.2 & 04:39:00.8 & $+$07:15:58 & 243 &	7.4$\pm$0.6 &	6.7$^{+1.0}_{-0.7}$ &	7.0$^{+1.3}_{-1.0}$ 	\\
A521\dotfill & 0.247 & 901 & 38.1 & 04:54:08.1 & $-$10:14:21 & 360 &	6.0$\pm$0.4 &	5.4$^{+0.5}_{-0.3}$ &	5.4$^{+0.4}_{-0.3}$ 	\\
A1835\dotfill & 0.252 & 495 & 18.4 & 14:01:01.9 & $+$02:52:41 & 187 &	8.2$\pm$0.2 &	16.3$^{+3.3}_{-2.5}$ &	16.1$^{+3.6}_{-2.9}$ 	\\
A68\dotfill & 0.255 & 3250 & 9.9 & 00:37:06.4 & $+$09:09:27 & 260 &	8.6$^{+1.4}_{-0.8}$ &	8.4$^{+1.9}_{-1.6}$ &	8.0$^{+2.0}_{-1.6}$ 	\\
MS 1455.0+2232\dotfill & 0.258 & 4192 & 91.6 & 14:57:15.1 & $+$22:20:34 & 148 &	4.7$\pm$0.1 &	5.6$\pm$0.3 &	5.6$\pm$0.3	\\
MS 1006.0+1202\dotfill & 0.261 & 925 & 15.4 & 10:08:47.5 & $+$11:47:34 & 234 &	6.1$\pm$0.4 &	6.6$^{+1.1}_{-0.7}$ &	6.6$^{+1.4}_{-0.7}$ 	\\
A697\dotfill & 0.282 & 4217 & 19.5 & 08:42:57.6 & $+$36:21:55 & 276 &	10.5$^{+0.9}_{-0.5}$ &	11.9$\pm$1.2 &	11.6$\pm$1.3	\\
A611\dotfill & 0.288 & 3194 & 24.3 & 08:00:56.8 & $+$36:03:23 & 172 &	8.9$^{+0.7}_{-0.6}$ &	11.8$^{+3.6}_{-2.2}$ &	12.5$^{+3.3}_{-2.8}$ 	\\
ZwCl 3146\dotfill & 0.291 & 909 & 43.7 & 10:23:39.6 & $+$04:11:10 & 246 &	6.5$\pm$0.1 &	8.7$^{+0.7}_{-0.4}$ &	8.6$^{+0.7}_{-0.5}$ 	\\
A781\dotfill & 0.298 & 534 & 9.9 & 09:20:21.6 & $+$30:30:20 & 264 &	5.3$^{+0.6}_{-0.4}$ &	5.3$^{+0.7}_{-0.4}$ &	5.2$^{+0.6}_{-0.4}$ 	\\
MS 1008.1-1224\dotfill & 0.301 & 926 & 28.6 & 10:10:32.2 & $-$12:39:23 & 196 &	6.4$\pm$0.4 &	6.5$^{+0.9}_{-0.6}$ &	6.6$^{+1.0}_{-0.6}$ 	\\
RXC J2245.0+2637\dotfill & 0.304 & 3287 & 14.6 & 22:45:04.9 & $+$26:38:02 & 150 &	5.9$\pm$0.3 &	7.1$^{+1.2}_{-0.9}$ &	6.7$^{+1.3}_{-0.8}$ 	\\
A1300\dotfill & 0.308 & 3276 & 13.7 & 11:31:55.3 & $-$19:54:46 & 268 &	8.8$^{+0.7}_{-0.6}$ &	9.4$^{+1.0}_{-0.9}$ &	9.1$^{+1.0}_{-0.9}$ 	\\
A2744\dotfill & 0.308 & 2212 & 22.1 & 00:14:15.3 & $-$30:22:50 & 235 &	10.1$\pm$0.6 &	9.2$^{+0.7}_{-0.6}$ &	9.3$\pm$0.7	\\
MS 2137.3-2353\dotfill & 0.313 & 5250 & 25.6 & 21:40:15.2 & $-$23:39:38 & 148 &	5.0$\pm$0.2 &	5.0$\pm$0.5 &	5.2$^{+0.3}_{-0.6}$ 	\\
A1995\dotfill & 0.318 & 906 & 10.0 & 14:52:58.6 & $+$58:02:58 & 191 &	8.1$^{+1.0}_{-0.8}$ &	6.0$^{+1.0}_{-0.8}$ &	5.7$^{+1.1}_{-0.8}$ 	\\
ZwCl 1358+6245\dotfill & 0.327 & 516 & 20.0 & 13:59:51.4 & $+$62:30:53 & 185 &	9.1$^{+0.9}_{-0.8}$ &	\nodata &	\nodata 	\\
A1722\dotfill & 0.328 & 3278 & 14.6 & 13:20:08.3 & $+$70:04:34 & 203 &	9.1$^{+1.5}_{-1.2}$ &	13.2$^{+6.4}_{-4.2}$ &	10.6$^{+7.4}_{-2.9}$ 	\\
RXC J0404.6+1109\dotfill & 0.355 & 3269 & 21.8 & 04:04:33.7 & $+$11:08:25 & 321 &	5.6$^{+0.8}_{-0.7}$ &	5.1$^{+0.9}_{-0.6}$ &	5.1$^{+1.0}_{-0.6}$ 	\\
RX J1532.9+3021\dotfill & 0.362 & 1649 & 8.1 & 15:32:54.0 & $+$30:21:04 & 128 &	6.1$\pm$0.3 &	8.1$^{+1.6}_{-1.2}$ &	7.5$^{+1.6}_{-1.1}$ 	\\
A370\dotfill & 0.373 & 515 & 53.9 & 02:39:54.5 & $-$01:34:47 & 184 &	8.7$^{+0.5}_{-0.4}$ &	8.1$\pm$0.5 &	7.8$\pm$0.5	\\
ZwCl 1953\dotfill & 0.374 & 1959 & 21.0 & 08:50:08.4 & $+$36:04:35 & 214 &	7.6$\pm$0.5 &	6.5$^{+0.6}_{-0.5}$ &	6.2$\pm$0.5	\\
RXC J0949.8+1707\dotfill & 0.383 & 3274 & 14.3 & 09:49:52.4 & $+$17:07:10 & 246 &	7.8$^{+0.7}_{-0.6}$ &	8.1$^{+1.2}_{-1.1}$ &	7.5$\pm$1.2	\\
ClG J1416+4446\dotfill & 0.400 & 541 & 29.9 & 14:16:28.4 & $+$44:46:42 & 128 &	3.8$\pm$0.3 &	4.5$^{+0.7}_{-0.5}$ &	4.3$^{+0.8}_{-0.5}$ 	\\
RXC J2228.6+2036\dotfill & 0.412 & 3285 & 19.8 & 22:28:32.1 & $+$20:37:23 & 244 &	8.1$\pm$0.5 &	7.9$^{+0.8}_{-0.7}$ &	8.4$^{+1.4}_{-0.8}$ 	\\
MS 0302.7+1658\dotfill & 0.426 & 525 & 10.0 & 03:05:31.7 & $+$17:10:05 & 82 &	3.6$^{+0.5}_{-0.4}$ &	2.8$^{+0.7}_{-0.5}$ &	2.7$^{+0.5}_{-0.4}$ 	\\
MS 1621.5+2640\dotfill & 0.426 & 546 & 30.0 & 16:23:35.0 & $+$26:34:26 & 197 &	6.4$^{+0.6}_{-0.5}$ &	6.4$^{+0.8}_{-0.7}$ &	6.3$^{+0.8}_{-0.7}$ 	\\
MACS J0417.5-1154\dotfill & 0.440 & 3270 & 11.9 & 04:17:33.5 & $-$11:53:58 & 270 &	9.4$\pm$0.7 &	11.4$^{+1.9}_{-1.6}$ &	10.6$^{+2.3}_{-1.3}$ 	\\
RXC J1206.2-0848\dotfill & 0.440 & 3277 & 23.4 & 12:06:12.2 & $-$08:48:05 & 236 &	11.4$\pm$0.9 &	12.5$^{+1.7}_{-1.5}$ &	12.7$^{+2.1}_{-1.8}$ 	\\
ClG J0329-0212\dotfill & 0.450 & 6108 & 39.5 & 03:29:41.6 & $-$02:11:46 & 127 &	5.9$\pm$0.2 &	6.8$^{+1.1}_{-0.6}$ &	7.2$^{+1.0}_{-0.8}$ 	\\
RX J1347.5-1145\dotfill & 0.451 & 3592 & 57.7 & 13:47:30.7 & $-$11:45:11 & 167 &	13.4$^{+0.5}_{-0.3}$ &	13.6$^{+1.7}_{-0.9}$ &	12.8$^{+1.3}_{-1.1}$ 	\\
ClG J1701+6414\dotfill & 0.453 & 547 & 48.2 & 17:01:24.0 & $+$64:14:11 & 108 &	4.7$\pm$0.3 &	5.3$^{+0.9}_{-0.5}$ &	5.1$^{+0.8}_{-0.6}$ 	\\
3C 295\dotfill & 0.461 & 2254 & 79.8 & 14:11:20.2 & $+$52:12:08 & 128 &	5.7$\pm$0.2 &	5.4$^{+0.6}_{-0.5}$ &	5.1$^{+0.7}_{-0.5}$ 	\\
ClG J1621+3810\dotfill & 0.461 & 6172 & 29.8 & 16:21:25.0 & $+$38:10:07 & 118 &	6.8$^{+0.6}_{-0.4}$ &	7.4$^{+1.4}_{-1.3}$ &	8.2$^{+2.4}_{-1.7}$ 	\\
ClG J1524+0957\dotfill & 0.516 & 1664 & 50.1 & 15:24:39.8 & $+$09:57:46 & 112 &	4.8$\pm$0.4 &	4.6$^{+0.6}_{-0.5}$ &	4.9$^{+0.7}_{-0.6}$ 	\\
MS 0451.6-0305\dotfill & 0.539 & 902 & 32.3 & 04:54:11.9 & $-$03:00:56 & 147 &	9.7$\pm$0.8 &	8.5$^{+1.1}_{-0.8}$ &	8.3$^{+1.4}_{-1.0}$ 	\\
MS 0015.9+1609\dotfill & 0.541 & 520 & 67.4 & 00:18:33.7 & $+$16:26:17 & 197 &	9.7$\pm$0.5 &	9.9$^{+0.7}_{-0.6}$ &	10.1$^{+0.9}_{-0.8}$ 	\\
ClG J1149+2223\dotfill & 0.544 & 3589 & 20.0 & 11:49:35.7 & $+$22:24:04 & 177 &	9.8$\pm$0.8 &	9.1$^{+1.0}_{-0.9}$ &	9.0$^{+1.2}_{-0.9}$ 	\\
ClG J1423+2404\dotfill & 0.545 & 4195 & 103.6 & 14:23:47.8 & $+$24:04:41 & 156 &	5.4$^{+0.2}_{-0.1}$ &	5.0$\pm$0.3 &	4.6$\pm$0.3	\\
ClG J1354-0221\dotfill & 0.546 & 5835 & 37.5 & 13:54:17.2 & $-$02:21:50 & 94 &	4.1$^{+0.8}_{-0.3}$ &	4.0$^{+1.1}_{-0.9}$ &	3.9$^{+1.2}_{-1.0}$ 	\\
ClG J0717+3745\dotfill & 0.548 & 4200 & 59.1 & 07:17:31.3 & $+$37:45:35 & 244 &	11.5$^{+0.7}_{-0.8}$ &	10.3$^{+0.8}_{-0.6}$ &	9.9$\pm$0.6	\\
ClG J1120+2326\dotfill & 0.562 & 1660 & 69.3 & 11:20:57.5 & $+$23:26:34 & 128 &	4.2$^{+0.6}_{-0.3}$ &	4.7$\pm$0.7 &	3.9$\pm$0.4	\\
ClG J2129-0741\dotfill & 0.570 & 3595 & 19.9 & 21:29:26.2 & $-$07:41:28 & 166 &	11.8$^{+2.8}_{-2.4}$ &	9.0$^{+2.7}_{-1.2}$ &	8.9$^{+3.5}_{-1.5}$ 	\\
MS 2053.7-0449\dotfill & 0.583 & 1667 & 44.5 & 20:56:21.3 & $-$04:37:49 & 69 &	4.0$^{+0.5}_{-0.2}$ &	3.6$^{+0.8}_{-0.5}$ &	3.2$^{+0.8}_{-0.5}$ 	\\
ClG J0647+7015\dotfill & 0.584 & 3584 & 19.9 & 06:47:50.6 & $+$70:14:54 & 160 &	15.0$^{+3.8}_{-2.7}$ &	\nodata &	\nodata 	\\
ClG J0542-4100\dotfill & 0.634 & 914 & 48.6 & 05:42:49.6 & $-$40:59:58 & 118 &	6.4$^{+0.8}_{-0.7}$ &	5.4$^{+1.0}_{-0.6}$ &	6.2$^{+1.2}_{-1.0}$ 	\\
ClG J1419+5326\dotfill & 0.640 & 3240 & 9.1 & 14:19:12.2 & $+$53:26:09 & 59 &	4.1$^{+0.8}_{-0.6}$ &	3.4$^{+0.8}_{-0.7}$ &	3.1$^{+1.6}_{-0.8}$ 	\\
ClG J0744+3927\dotfill & 0.686 & 6111 & 49.5 & 07:44:52.8 & $+$39:27:27 & 118 &	9.6$\pm$0.9 &	11.7$^{+2.2}_{-2.0}$ &	10.4$^{+3.2}_{-2.1}$ 	\\
ClG J1221+4918\dotfill & 0.700 & 1662 & 78.3 & 12:21:25.9 & $+$49:18:28 & 138 &	6.5$^{+0.8}_{-0.6}$ &	6.4$^{+1.0}_{-0.7}$ &	6.1$^{+0.9}_{-0.8}$ 	\\
ClG J1113-2615\dotfill & 0.730 & 915 & 62.5 & 11:13:05.0 & $-$26:15:40 & 79 &	3.7$^{+0.6}_{-0.5}$ &	2.8$^{+0.6}_{-0.4}$ &	2.6$^{+0.6}_{-0.4}$ 	\\
ClG 1137+6625\dotfill & 0.782 & 536 & 27.6 & 11:40:22.4 & $+$66:08:16 & 98 &	5.9$^{+1.2}_{-0.9}$ &	6.1$^{+2.5}_{-1.7}$ &	6.4$^{+4.4}_{-2.2}$ 	\\
RX J1350.0+6007\dotfill & 0.804 & 2229 & 58.3 & 13:50:48.3 & $+$60:07:11 & 98 &	4.1$^{+0.8}_{-0.6}$ &	4.3$^{+1.6}_{-0.8}$ &	4.5$^{+2.1}_{-1.2}$ 	\\
RX J1317+2911\dotfill & 0.805 & 2228 & 111.3 & 13:17:21.8 & $+$29:11:19 & 69 &	3.8$^{+1.7}_{-0.9}$ &	3.3$^{+3.1}_{-1.1}$ &	2.2$^{+3.0}_{-0.5}$ 	\\
RX J1716+6708\dotfill & 0.813 & 548 & 51.2 & 17:16:49.1 & $+$67:08:24 & 108 &	6.4$^{+0.9}_{-0.8}$ &	5.6$^{+1.2}_{-0.8}$ &	6.4$^{+2.3}_{-1.4}$ 	\\
ClG J1056-0337\dotfill & 0.826 & 512 & 66.7 & 10:56:59.5 & $-$03:37:34 & 118 &	9.2$^{+1.5}_{-1.2}$ &	8.7$^{+1.8}_{-1.1}$ &	8.6$^{+2.3}_{-1.5}$ 	\\
ClG J1226+3332\dotfill & 0.890 & 3180 & 31.6 & 12:26:58.0 & $+$33:32:46 & 108 &	12.2$^{+1.8}_{-1.7}$ &	13.6$^{+4.0}_{-3.2}$ &	10.3$^{+5.1}_{-3.1}$ 	\\
ClG J1415+3611\dotfill & 1.030 & 4163 & 89.2 & 14:15:11.2 & $+$36:12:03 & 79 &	6.8$^{+1.0}_{-0.7}$ &	6.2$^{+1.8}_{-1.1}$ &	6.0$^{+1.5}_{-1.6}$ 	\\
ClG J1252-2927\dotfill & 1.235 & 4198 & 162.5 & 12:52:54.4 & $-$29:27:16 & 69 &	5.7$^{+1.4}_{-1.0}$ &	5.3$^{+1.6}_{-1.0}$ &	5.2$^{+2.4}_{-1.3}$ 	\\
\enddata
\tablenotetext{a}{Exposure time after light curve filtering.}
\tablenotetext{b}{Coordinates given are center of spectral extraction aperture.}
\tablenotetext{c}{Core-subtracted temperature measured assuming self-similar evolution of \rd.}
\tablenotetext{d}{Core-subtracted temperature measured assuming no evolution of \rd.}
\label{tab:basicinfo}
\end{deluxetable*}

\tabletypesize{\tiny}
\begin{deluxetable*}{p{2.2cm}cccccc}

\tablewidth{0pt}
\tabletypesize{\tiny}
\tablecaption{$\beta$ Model Parameters}

\tablehead{
\colhead{Cluster} & \colhead{Fit aperture} &
\colhead{$\beta$} & \colhead{$I_1$} & \colhead{$R_{c,1}$} & 
\colhead{$I_2$} & \colhead{$R_{c,2}$} \\ 
 & \colhead{(arcsec)} & & \colhead{(erg s$^{-1}$ cm$^{-2}$ arcmin$^{-2}$)} & 
 \colhead{(arcsec)} & \colhead{(erg s$^{-1}$ cm$^{-2}$ arcmin$^{-2}$)} & \colhead{(arcssec)} 
}
\startdata
A665\dotfill  & 394 &	0.62$\pm$0.01 &	1.9$\pm$0.0 E-12 &	65.6$^{+1.4}_{-1.3}$ &	\nodata &	\nodata	\\
A963\dotfill  & 197 &	0.55$\pm$0.00 &	6.6$\pm$0.1 E-12 &	21.1$^{+0.5}_{-0.4}$ &	\nodata &	\nodata	\\
RX J0439.0+0520\dotfill  & 148 &	0.67$^{+0.04}_{-0.02}$ &	3.0$^{+0.5}_{-0.4}$E-12 &	28.0$^{+3.7}_{-3.3}$ &	5.4$\pm$0.4 E-11 &	5.4$^{+0.5}_{-0.4}$ 	\\
A1423\dotfill  & 187 &	0.46$\pm$0.01 &	7.2$^{+0.6}_{-0.5}$E-12 &	10.5$\pm$0.8 &	\nodata &	\nodata	\\
ZwCl 2701\dotfill  & 153 &	0.58$\pm$0.01 &	1.5$\pm$0.0 E-11 &	12.3$\pm$0.3 &	\nodata &	\nodata	\\
A773\dotfill  & 256 &	0.60$\pm$0.01 &	2.6$\pm$0.1 E-12 &	41.2$^{+1.4}_{-1.5}$ &	\nodata &	\nodata	\\
A2261\dotfill  & 148 &	0.55$^{+0.01}_{-0.00}$ &	1.2$\pm$0.0 E-11 &	18.1$^{+0.6}_{-0.5}$ &	\nodata &	\nodata	\\
ACO 2246\dotfill  & 148 &	0.52$\pm$0.01 &	3.7$\pm$0.3 E-12 &	9.0$^{+0.8}_{-0.7}$ &	\nodata &	\nodata	\\
A1682\dotfill  & 177 &	0.56$^{+0.06}_{-0.04}$ &	1.0$\pm$0.1 E-12 &	49.2$^{+9.7}_{-7.5}$ &	\nodata &	\nodata	\\
A2111\dotfill  & 295 &	0.58$\pm$0.02 &	1.2$\pm$0.1 E-12 &	48.7$^{+3.1}_{-3.2}$ &	\nodata &	\nodata	\\
A267\dotfill  & 276 &	0.62$\pm$0.01 &	3.3$\pm$0.1 E-12 &	33.3$\pm$1.2 &	\nodata &	\nodata	\\
RX J2129.7+0005\dotfill  & 157 &	0.60$\pm$0.01 &	7.4$\pm$0.9 E-12 &	23.4$^{+2.3}_{-2.0}$ &	6.2$^{+0.6}_{-0.3}$E-11 &	4.1$\pm$0.4 	\\
RX J0439.0+0715\dotfill  & 256 &	0.61$\pm$0.01 &	6.0$^{+0.2}_{-0.1}$E-12 &	26.1$^{+0.8}_{-1.1}$ &	\nodata &	\nodata	\\
A521\dotfill  & 295 &	0.75$\pm$0.00 &	5.5$\pm$0.2 E-13 &	122.0$\pm$2.4 &	\nodata &	\nodata	\\
A1835\dotfill  & 167 &	0.73$\pm$0.01 &	5.7$\pm$0.2 E-12 &	44.8$^{+1.2}_{-1.4}$ &	1.1$\pm$0.0 E-10 &	8.9$\pm$0.2 	\\
A68\dotfill  & 246 &	0.75$^{+0.03}_{-0.02}$ &	2.3$\pm$0.1 E-12 &	53.0$^{+3.3}_{-3.0}$ &	\nodata &	\nodata	\\
MS 1455.0+2232\dotfill  & 148 &	0.61$\pm$0.00 &	6.4$\pm$0.1 E-11 &	8.9$\pm$0.1 &	\nodata &	\nodata	\\
MS 1006.0+1202\dotfill  & 216 &	0.70$\pm$0.02 &	1.7$\pm$0.1 E-12 &	48.0$^{+2.8}_{-2.5}$ &	\nodata &	\nodata	\\
A697\dotfill  & 256 &	0.64$\pm$0.01 &	3.7$\pm$0.1 E-12 &	46.6$^{+1.6}_{-1.5}$ &	\nodata &	\nodata	\\
A611\dotfill  & 172 &	0.60$\pm$0.01 &	8.4$\pm$0.3 E-12 &	18.4$\pm$0.6 &	\nodata &	\nodata	\\
ZwCl 3146\dotfill  & 246 &	0.68$^{+0.01}_{-0.00}$ &	1.6$\pm$0.1 E-11 &	23.8$\pm$0.6 &	1.2$\pm$0.0 E-10 &	5.4$\pm$0.1 	\\
A781\dotfill  & 226 &	1.47$^{+0.36}_{-0.24}$ &	6.9$\pm$0.3 E-13 &	157.2$^{+28.0}_{-21.2}$ &	\nodata &	\nodata	\\
MS 1008.1-1224\dotfill  & 197 &	0.65$\pm$0.02 &	2.4$\pm$0.1 E-12 &	35.0$\pm$1.8 &	\nodata &	\nodata	\\
RXC J2245.0+2637\dotfill  & 148 &	0.66$\pm$0.02 &	8.5$\pm$0.8 E-12 &	21.3$^{+2.2}_{-1.6}$ &	4.1$^{+0.6}_{-0.5}$E-11 &	4.1$^{+0.6}_{-0.5}$ 	\\
A1300\dotfill  & 207 &	0.49$\pm$0.01 &	4.9$^{+0.3}_{-0.2}$E-12 &	22.1$^{+1.2}_{-1.3}$ &	\nodata &	\nodata	\\
A2744\dotfill  & 406 &	1.10$\pm$0.04 &	1.8$\pm$0.0 E-12 &	112.5$^{+4.3}_{-3.8}$ &	\nodata &	\nodata	\\
MS 2137.3-2353\dotfill  & 138 &	0.64$\pm$0.01 &	3.9$^{+1.2}_{-0.9}$E-11 &	10.2$\pm$1.1 &	1.1$\pm$0.0 E-10 &	3.3$^{+0.5}_{-0.6}$ 	\\
A1995\dotfill  & 216 &	0.82$\pm$0.03 &	3.5$\pm$0.1 E-12 &	43.5$^{+2.2}_{-2.1}$ &	\nodata &	\nodata	\\
ZwCl 1358+6245\dotfill  & 157 &	0.66$^{+0.03}_{-0.01}$ &	2.8$\pm$0.2 E-12 &	31.4$^{+2.4}_{-1.8}$ &	4.0$^{+0.2}_{-0.4}$E-11 &	3.8$^{+0.4}_{-0.2}$ 	\\
A1722\dotfill  & 148 &	0.64$^{+0.05}_{-0.03}$ &	2.1$^{+0.4}_{-0.6}$E-12 &	30.7$^{+7.9}_{-3.9}$ &	2.9$^{+1.2}_{-0.8}$E-12 &	7.2$^{+4.4}_{-2.5}$ 	\\
RXC J0404.6+1109\dotfill  & 128 &	0.46$^{+0.04}_{-0.03}$ &	1.0$\pm$0.1 E-12 &	28.6$^{+5.9}_{-4.9}$ &	\nodata &	\nodata	\\
RX J1532.9+3021\dotfill  & 118 &	0.61$\pm$0.01 &	1.1$\pm$0.0 E-10 &	7.8$^{+0.2}_{-0.3}$ &	\nodata &	\nodata	\\
A370\dotfill  & 187 &	0.81$\pm$0.02 &	1.7$\pm$0.0 E-12 &	59.4$^{+2.1}_{-2.0}$ &	\nodata &	\nodata	\\
ZwCl 1953\dotfill  & 246 &	0.65$\pm$0.01 &	4.5$\pm$0.2 E-12 &	30.9$^{+1.3}_{-1.4}$ &	\nodata &	\nodata	\\
RXC J0949.8+1707\dotfill  & 153 &	0.63$\pm$0.02 &	6.0$\pm$0.3 E-12 &	27.8$^{+1.9}_{-1.8}$ &	\nodata &	\nodata	\\
ClG J1416+4446\dotfill  & 148 &	0.58$^{+0.03}_{-0.02}$ &	3.0$^{+0.5}_{-0.7}$E-12 &	16.6$^{+3.3}_{-2.4}$ &	2.3$^{+0.9}_{-0.6}$E-11 &	1.9$^{+0.8}_{-0.5}$ 	\\
RXC J2228.6+2036\dotfill  & 172 &	0.64$\pm$0.02 &	5.3$\pm$0.2 E-12 &	31.0$^{+1.8}_{-1.7}$ &	\nodata &	\nodata	\\
MS 0302.7+1658\dotfill  & 98 &	0.54$\pm$0.02 &	1.5$^{+0.3}_{-0.2}$E-11 &	6.9$^{+1.2}_{-1.0}$ &	\nodata &	\nodata	\\
MS 1621.5+2640\dotfill  & 197 &	0.67$\pm$0.03 &	1.3$\pm$0.1 E-12 &	41.6$^{+3.6}_{-2.9}$ &	\nodata &	\nodata	\\
MACS J0417.5-1154\dotfill  & 192 &	0.65$\pm$0.02 &	4.4$\pm$0.3 E-12 &	48.4$^{+2.8}_{-3.3}$ &	1.1$\pm$0.1 E-10 &	5.2$\pm$0.3 	\\
RXC J1206.2-0848\dotfill  & 197 &	0.60$\pm$0.01 &	1.7$\pm$0.1 E-11 &	19.1$^{+0.7}_{-0.6}$ &	\nodata &	\nodata	\\
ClG J0329-0212\dotfill  & 118 &	0.52$\pm$0.00 &	1.2$\pm$0.1 E-10 &	3.8$^{+0.1}_{-0.2}$ &	\nodata &	\nodata	\\
RX J1347.5-1145\dotfill  & 189 &	0.65$\pm$0.00 &	4.5$\pm$0.2 E-11 &	16.7$^{+0.4}_{-0.5}$ &	3.9$\pm$0.1 E-10 &	3.7$\pm$0.1 	\\
ClG J1701+6414\dotfill  & 148 &	0.58$\pm$0.02 &	1.1$^{+0.2}_{-0.1}$E-12 &	29.4$^{+3.1}_{-3.4}$ &	2.9$^{+0.5}_{-0.3}$E-11 &	2.7$\pm$0.3 	\\
3C 295\dotfill  & 118 &	0.63$\pm$0.01 &	9.3$^{+1.7}_{-1.4}$E-12 &	13.1$^{+1.4}_{-1.3}$ &	1.3$\pm$0.1 E-10 &	2.7$^{+0.2}_{-0.1}$ 	\\
ClG J1621+3810\dotfill  & 108 &	0.60$\pm$0.02 &	7.6$^{+2.0}_{-1.6}$E-12 &	14.8$^{+2.5}_{-2.0}$ &	9.2$^{+1.2}_{-1.0}$E-11 &	2.5$^{+0.4}_{-0.3}$ 	\\
ClG J1524+0957\dotfill  & 128 &	0.95$^{+0.14}_{-0.10}$ &	1.1$\pm$0.1 E-12 &	56.2$^{+8.2}_{-6.6}$ &	\nodata &	\nodata	\\
MS 0451.6-0305\dotfill  & 459 &	0.85$\pm$0.02 &	9.2$\pm$0.2 E-12 &	37.9$\pm$1.1 &	\nodata &	\nodata	\\
MS 0015.9+1609\dotfill  & 216 &	0.70$\pm$0.01 &	6.5$\pm$0.2 E-12 &	37.5$^{+1.1}_{-1.0}$ &	\nodata &	\nodata	\\
ClG J1149+2223\dotfill  & 295 &	0.65$\pm$0.02 &	4.3$\pm$0.2 E-12 &	40.9$^{+2.5}_{-2.2}$ &	\nodata &	\nodata	\\
ClG J1423+2404\dotfill  & 98 &	0.65$\pm$0.01 &	4.2$\pm$0.4 E-12 &	22.0$^{+1.4}_{-1.0}$ &	2.3$\pm$0.0 E-10 &	3.7$\pm$0.1 	\\
ClG J1354-0221\dotfill  & 157 &	0.76$^{+0.12}_{-0.08}$ &	8.3$^{+1.0}_{-0.8}$E-13 &	39.8$^{+8.3}_{-6.6}$ &	\nodata &	\nodata	\\
ClG J0717+3745\dotfill  & 187 &	0.82$\pm$0.02 &	4.7$\pm$0.1 E-12 &	65.6$^{+2.5}_{-2.2}$ &	\nodata &	\nodata	\\
ClG J1120+2326\dotfill  & 148 &	1.74$^{+0.54}_{-0.31}$ &	9.9$^{+0.5}_{-0.4}$E-13 &	88.4$^{+18.5}_{-12.6}$ &	\nodata &	\nodata	\\
ClG J2129-0741\dotfill  & 166 &	0.62$^{+0.02}_{-0.01}$ &	1.1$\pm$0.1 E-11 &	18.6$\pm$1.2 &	\nodata &	\nodata	\\
MS 2053.7-0449\dotfill  & 89 &	0.63$^{+0.05}_{-0.04}$ &	3.9$^{+0.5}_{-0.4}$E-12 &	15.6$^{+2.5}_{-2.1}$ &	\nodata &	\nodata	\\
ClG J0647+7015\dotfill  & 148 &	0.63$\pm$0.02 &	1.3$\pm$0.1 E-11 &	18.4$^{+1.3}_{-1.2}$ &	\nodata &	\nodata	\\
ClG J0542-4100\dotfill  & 112 &	0.58$\pm$0.03 &	2.9$\pm$0.2 E-12 &	22.5$^{+2.7}_{-2.4}$ &	\nodata &	\nodata	\\
ClG J1419+5326\dotfill  & 118 &	0.60$\pm$0.03 &	1.9$^{+0.4}_{-0.3}$E-11 &	7.3$^{+1.2}_{-1.1}$ &	\nodata &	\nodata	\\
ClG J0744+3927\dotfill  & 148 &	0.56$\pm$0.01 &	4.4$\pm$0.2 E-11 &	8.5$\pm$0.4 &	\nodata &	\nodata	\\
ClG J1221+4918\dotfill  & 157 &	0.73$^{+0.04}_{-0.03}$ &	2.5$\pm$0.1 E-12 &	35.7$^{+2.8}_{-2.6}$ &	\nodata &	\nodata	\\
ClG J1113-2615\dotfill  & 98 &	0.73$^{+0.08}_{-0.06}$ &	4.3$\pm$0.5 E-12 &	15.8$^{+2.8}_{-2.5}$ &	\nodata &	\nodata	\\
ClG 1137+6625\dotfill  & 98 &	0.65$^{+0.04}_{-0.03}$ &	1.5$^{+0.2}_{-0.1}$E-11 &	12.4$^{+1.4}_{-1.3}$ &	\nodata &	\nodata	\\
RX J1350.0+6007\dotfill  & 148 &	0.61$^{+0.05}_{-0.04}$ &	2.3$\pm$0.3 E-12 &	21.4$^{+3.9}_{-3.1}$ &	\nodata &	\nodata	\\
RX J1317+2911\dotfill  & 89 &	0.84$^{+0.40}_{-0.20}$ &	5.3$^{+1.9}_{-1.2}$E-13 &	29.4$^{+13.8}_{-9.8}$ &	4.8$^{+2.1}_{-1.2}$E-12 &	4.3$^{+2.4}_{-1.6}$ 	\\
RX J1716+6708\dotfill  & 148 &	0.68$^{+0.04}_{-0.03}$ &	8.3$\pm$0.6 E-12 &	17.3$^{+1.9}_{-1.7}$ &	\nodata &	\nodata	\\
ClG J1056-0337\dotfill  & 197 &	0.67$\pm$0.00 &	5.8$\pm$0.3 E-12 &	31.9$\pm$0.9 &	\nodata &	\nodata	\\
ClG J1226+3332\dotfill  & 128 &	0.68$\pm$0.02 &	3.3$\pm$0.2 E-11 &	14.5$\pm$1.0 &	\nodata &	\nodata	\\
ClG J1415+3611\dotfill  & 98 &	0.75$^{+0.06}_{-0.04}$ &	9.5$\pm$1.2 E-12 &	18.1$\pm$2.4 &	6.5$^{+2.0}_{-1.4}$E-11 &	2.5$^{+0.6}_{-0.5}$ 	\\
ClG J1252-2927\dotfill  & 89 &	0.54$\pm$0.03 &	1.1$\pm$0.2 E-11 &	8.8$^{+1.6}_{-1.4}$ &	\nodata &	\nodata	\\
\enddata
\label{tab:betainfo}
\end{deluxetable*}

\tabletypesize{\scriptsize}
\begin{deluxetable*}{p{2.5cm}ccccccc}

\tablewidth{0pt}
\tabletypesize{\tiny}
\tablecaption{Cluster Measurements Assuming Self-Similar Evolution}
\tablehead{
Cluster &
\colhead{$L_{\rm X,2500}$} & \colhead{$L_{\rm X,500}$} &
\colhead{$M_{\rm g,2500}$} & \colhead{$M_{\rm g,500}$} &
\colhead{$R_{1.5{\rm E}-13}$} & \colhead{$R_{6{\rm E}-14}$} & \colhead{$R_{3{\rm E}-14}$} \\
&  \colhead{($10^{44}L_\odot$)} & \colhead{($10^{44}L_\odot$)} &
\colhead{($10^{13}M_\odot$)} & \colhead{($10^{13}M_\odot$)} &
\colhead{(Mpc)} & \colhead{(Mpc)} & \colhead{(Mpc)}
}
\startdata
A665\dotfill & 		3.27$\pm$0.03 &	\nodata &	3.93$\pm$0.09 &	13.05$\pm$0.30 &	0.40$\pm$0.02 & 	\nodata &	\nodata \\
A963\dotfill & 		2.95$\pm$0.03 &	\nodata &	2.86$\pm$0.08 &	9.27$\pm$0.27 &	0.33$\pm$0.02 & 	\nodata &	\nodata \\
RX J0439.0+0520\dotfill & 		2.17$\pm$0.02 &	\nodata &	1.42$\pm$0.07 &	3.79$\pm$0.19 &	0.23$\pm$0.01 & 	0.31$\pm$0.02 & 	0.40$\pm$0.04 \\ 
A1423\dotfill & 		1.98$\pm$0.04 &	\nodata &	2.08$\pm$0.07 &	8.01$\pm$0.28 &	0.28$\pm$0.01 & 	0.46$\pm$0.04 & 	\nodata \\
ZwCl 2701\dotfill & 		2.10$\pm$0.01 &	\nodata &	1.58$\pm$0.07 &	4.63$\pm$0.21 &	0.24$\pm$0.01 & 	0.34$\pm$0.02 & 	\nodata \\
A773\dotfill & 		2.92$\pm$0.03 &	\nodata &	3.55$\pm$0.09 &	11.34$\pm$0.30 &	0.36$\pm$0.02 & 	\nodata &	\nodata \\
A2261\dotfill & 		4.69$\pm$0.03 &	\nodata &	3.81$\pm$0.11 &	12.20$\pm$0.36 &	0.38$\pm$0.02 & 	\nodata &	\nodata \\
ACO 2246\dotfill & 		0.46$\pm$0.01 &	\nodata &	0.56$\pm$0.04 &	1.91$\pm$0.15 &	\nodata &	0.21$\pm$0.01 & 	\nodata \\
A1682\dotfill & 		1.61$\pm$0.09 &	\nodata &	2.01$\pm$0.08 &	7.55$\pm$0.31 &	0.32$\pm$0.02 & 	0.52$\pm$0.05 & 	\nodata \\
A2111\dotfill & 		1.77$\pm$0.05 &	\nodata &	2.59$\pm$0.08 &	9.03$\pm$0.29 &	0.31$\pm$0.01 & 	0.49$\pm$0.03 & 	0.66$\pm$0.06 \\ 
A267\dotfill & 		2.37$\pm$0.03 &	\nodata &	2.73$\pm$0.09 &	8.15$\pm$0.27 &	0.30$\pm$0.01 & 	0.43$\pm$0.03 & 	0.57$\pm$0.05 \\ 
RX J2129.7+0005\dotfill & 		4.26$\pm$0.03 &	\nodata &	2.75$\pm$0.11 &	8.13$\pm$0.33 &	0.34$\pm$0.02 & 	0.49$\pm$0.04 & 	0.63$\pm$0.07 \\ 
RX J0439.0+0715\dotfill & 		3.28$\pm$0.04 &	\nodata &	3.20$\pm$0.11 &	9.43$\pm$0.31 &	0.35$\pm$0.02 & 	0.49$\pm$0.03 & 	0.61$\pm$0.05 \\ 
A521\dotfill & 		2.00$\pm$0.07 &	3.43$\pm$0.07 &	2.38$\pm$0.10 &	10.80$\pm$0.44 &	0.43$\pm$0.03 & 	0.68$\pm$0.08 & 	0.82$\pm$0.11 \\ 
A1835\dotfill & 		10.50$\pm$0.05 &	\nodata &	5.35$\pm$0.16 &	13.56$\pm$0.42 &	0.44$\pm$0.02 & 	\nodata &	\nodata \\
A68\dotfill & 		2.84$\pm$0.07 &	\nodata &	3.68$\pm$0.11 &	9.96$\pm$0.29 &	0.35$\pm$0.01 & 	0.48$\pm$0.03 & 	\nodata \\
MS 1455.0+2232\dotfill & 		4.95$\pm$0.01 &	5.45$\pm$0.04 &	2.20$\pm$0.12 &	6.05$\pm$0.33 &	0.30$\pm$0.02 & 	0.41$\pm$0.04 & 	\nodata \\
MS 1006.0+1202\dotfill & 		1.85$\pm$0.03 &	\nodata &	2.37$\pm$0.10 &	7.21$\pm$0.31 &	0.29$\pm$0.01 & 	0.41$\pm$0.03 & 	\nodata \\
A697\dotfill & 		5.32$\pm$0.09 &	\nodata &	5.96$\pm$0.16 &	18.56$\pm$0.50 &	0.51$\pm$0.02 & 	0.71$\pm$0.05 & 	0.88$\pm$0.07 \\ 
A611\dotfill & 		2.91$\pm$0.03 &	\nodata &	3.29$\pm$0.11 &	9.50$\pm$0.31 &	0.32$\pm$0.01 & 	0.44$\pm$0.02 & 	\nodata \\
ZwCl 3146\dotfill & 		9.16$\pm$0.02 &	\nodata &	4.19$\pm$0.19 &	10.77$\pm$0.48 &	0.41$\pm$0.03 & 	0.56$\pm$0.05 & 	\nodata \\
A781\dotfill & 		1.43$\pm$0.09 &	\nodata &	2.05$\pm$0.12 &	7.99$\pm$0.45 &	0.37$\pm$0.03 & 	0.59$\pm$0.07 & 	0.74$\pm$0.10 \\ 
MS 1008.1-1224\dotfill & 		1.93$\pm$0.03 &	\nodata &	2.44$\pm$0.11 &	7.44$\pm$0.35 &	0.33$\pm$0.02 & 	0.45$\pm$0.03 & 	0.56$\pm$0.05 \\ 
RXC J2245.0+2637\dotfill & 		3.47$\pm$0.02 &	3.85$\pm$0.05 &	2.54$\pm$0.13 &	6.82$\pm$0.35 &	0.30$\pm$0.02 & 	0.41$\pm$0.03 & 	0.51$\pm$0.04 \\ 
A1300\dotfill & 		4.04$\pm$0.09 &	5.55$\pm$0.11 &	4.28$\pm$0.15 &	15.90$\pm$0.56 &	0.46$\pm$0.02 & 	0.69$\pm$0.05 & 	\nodata \\
A2744\dotfill & 		4.74$\pm$0.10 &	\nodata &	6.10$\pm$0.19 &	17.12$\pm$0.52 &	0.58$\pm$0.03 & 	\nodata &	\nodata \\
MS 2137.3-2353\dotfill & 		5.15$\pm$0.02 &	\nodata &	2.27$\pm$0.14 &	5.87$\pm$0.37 &	0.29$\pm$0.02 & 	0.39$\pm$0.03 & 	0.49$\pm$0.05 \\ 
A1995\dotfill & 		2.88$\pm$0.04 &	\nodata &	3.45$\pm$0.14 &	8.43$\pm$0.33 &	0.34$\pm$0.01 & 	0.44$\pm$0.02 & 	0.56$\pm$0.04 \\ 
ZwCl 1358+6245\dotfill & 		2.64$\pm$0.04 &	\nodata &	3.28$\pm$0.12 &	9.21$\pm$0.33 &	0.31$\pm$0.01 & 	0.43$\pm$0.02 & 	0.54$\pm$0.03 \\ 
A1722\dotfill & 		1.80$\pm$0.05 &	\nodata &	2.90$\pm$0.10 &	8.53$\pm$0.31 &	0.28$\pm$0.01 & 	0.40$\pm$0.02 & 	0.54$\pm$0.03 \\ 
RXC J0404.6+1109\dotfill & 		1.20$\pm$0.08 &	2.30$\pm$0.14 &	1.73$\pm$0.11 &	7.61$\pm$0.48 &	0.28$\pm$0.01 & 	0.64$\pm$0.07 & 	\nodata \\
RX J1532.9+3021\dotfill & 		8.84$\pm$0.03 &	\nodata &	3.62$\pm$0.22 &	9.81$\pm$0.59 &	0.36$\pm$0.02 & 	0.50$\pm$0.04 & 	0.64$\pm$0.07 \\ 
A370\dotfill & 		2.59$\pm$0.05 &	\nodata &	3.84$\pm$0.16 &	11.46$\pm$0.49 &	0.40$\pm$0.02 & 	\nodata &	\nodata \\
ZwCl 1953\dotfill & 		3.43$\pm$0.06 &	4.31$\pm$0.09 &	3.68$\pm$0.18 &	11.27$\pm$0.56 &	0.39$\pm$0.02 & 	0.56$\pm$0.04 & 	0.71$\pm$0.07 \\ 
RXC J0949.8+1707\dotfill & 		4.08$\pm$0.07 &	5.11$\pm$0.09 &	4.05$\pm$0.20 &	12.40$\pm$0.61 &	0.41$\pm$0.02 & 	0.57$\pm$0.04 & 	0.76$\pm$0.07 \\ 
ClG J1416+4446\dotfill & 		1.10$\pm$0.02 &	1.44$\pm$0.05 &	1.14$\pm$0.12 &	3.78$\pm$0.40 &	0.24$\pm$0.01 & 	0.34$\pm$0.03 & 	0.44$\pm$0.05 \\ 
RXC J2228.6+2036\dotfill & 		4.15$\pm$0.08 &	5.66$\pm$0.08 &	4.43$\pm$0.23 &	13.92$\pm$0.71 &	0.50$\pm$0.03 & 	0.69$\pm$0.06 & 	0.82$\pm$0.08 \\ 
MS 0302.7+1658\dotfill & 		1.41$\pm$0.02 &	\nodata &	1.15$\pm$0.14 &	3.76$\pm$0.45 &	0.23$\pm$0.01 & 	0.33$\pm$0.03 & 	0.43$\pm$0.05 \\ 
MS 1621.5+2640\dotfill & 		1.41$\pm$0.05 &	2.09$\pm$0.09 &	2.15$\pm$0.14 &	7.60$\pm$0.51 &	0.32$\pm$0.02 & 	0.48$\pm$0.04 & 	0.70$\pm$0.08 \\ 
MACS J0417.5-1154\dotfill & 		10.27$\pm$0.19 &	13.17$\pm$0.11 &	7.14$\pm$0.33 &	24.15$\pm$1.13 &	0.63$\pm$0.04 & 	0.82$\pm$0.07 & 	0.97$\pm$0.10 \\ 
RXC J1206.2-0848\dotfill & 		8.02$\pm$0.10 &	9.47$\pm$0.11 &	7.24$\pm$0.28 &	21.76$\pm$0.84 &	0.54$\pm$0.03 & 	0.73$\pm$0.05 & 	0.92$\pm$0.07 \\ 
ClG J0329-0212\dotfill & 		5.23$\pm$0.03 &	5.92$\pm$0.09 &	2.89$\pm$0.22 &	9.37$\pm$0.72 &	0.37$\pm$0.02 & 	0.52$\pm$0.05 & 	0.65$\pm$0.07 \\ 
RX J1347.5-1145\dotfill & 		18.85$\pm$0.05 &	\nodata &	10.10$\pm$0.34 &	25.94$\pm$0.87 &	0.56$\pm$0.02 & 	\nodata &	\nodata \\
ClG J1701+6414\dotfill & 		1.22$\pm$0.03 &	\nodata &	1.41$\pm$0.14 &	5.16$\pm$0.50 &	0.26$\pm$0.01 & 	0.38$\pm$0.03 & 	0.50$\pm$0.05 \\ 
3C 295\dotfill & 		2.91$\pm$0.01 &	\nodata &	2.06$\pm$0.17 &	5.62$\pm$0.45 &	0.27$\pm$0.01 & 	0.36$\pm$0.02 & 	0.46$\pm$0.04 \\ 
ClG J1621+3810\dotfill & 		3.09$\pm$0.03 &	\nodata &	2.68$\pm$0.18 &	7.99$\pm$0.54 &	0.31$\pm$0.01 & 	0.46$\pm$0.03 & 	0.59$\pm$0.05 \\ 
ClG J1524+0957\dotfill & 		0.92$\pm$0.04 &	1.47$\pm$0.08 &	1.47$\pm$0.16 &	5.28$\pm$0.57 &	0.29$\pm$0.02 & 	0.43$\pm$0.04 & 	0.52$\pm$0.06 \\ 
MS 0451.6-0305\dotfill & 		6.09$\pm$0.09 &	7.06$\pm$0.16 &	6.18$\pm$0.34 &	15.86$\pm$0.88 &	0.48$\pm$0.02 & 	0.62$\pm$0.04 & 	\nodata \\
MS 0015.9+1609\dotfill & 		5.76$\pm$0.08 &	7.61$\pm$0.11 &	6.14$\pm$0.34 &	19.43$\pm$1.08 &	0.55$\pm$0.03 & 	\nodata &	\nodata \\
ClG J1149+2223\dotfill & 		4.78$\pm$0.14 &	7.07$\pm$0.14 &	5.66$\pm$0.32 &	20.17$\pm$1.12 &	0.56$\pm$0.03 & 	0.86$\pm$0.08 & 	1.08$\pm$0.12 \\ 
ClG J1423+2404\dotfill & 		5.54$\pm$0.02 &	6.23$\pm$0.10 &	2.75$\pm$0.28 &	7.91$\pm$0.80 &	0.34$\pm$0.02 & 	0.49$\pm$0.05 & 	\nodata \\
ClG J1354-0221\dotfill & 		0.53$\pm$0.03 &	\nodata &	0.94$\pm$0.13 &	3.58$\pm$0.48 &	0.20$\pm$0.01 & 	0.33$\pm$0.03 & 	0.44$\pm$0.05 \\ 
ClG J0717+3745\dotfill & 		7.78$\pm$0.23 &	11.13$\pm$0.13 &	8.51$\pm$0.41 &	29.57$\pm$1.41 &	0.68$\pm$0.04 & 	0.88$\pm$0.07 & 	\nodata \\
ClG J1120+2326\dotfill & 		0.78$\pm$0.05 &	\nodata &	1.22$\pm$0.16 &	4.41$\pm$0.58 &	0.26$\pm$0.02 & 	0.37$\pm$0.03 & 	0.43$\pm$0.04 \\ 
ClG J2129-0741\dotfill & 		4.22$\pm$0.17 &	\nodata &	5.56$\pm$0.27 &	16.45$\pm$0.80 &	0.42$\pm$0.01 & 	0.63$\pm$0.03 & 	0.80$\pm$0.05 \\ 
MS 2053.7-0449\dotfill & 		0.89$\pm$0.02 &	\nodata &	1.13$\pm$0.17 &	3.58$\pm$0.52 &	0.23$\pm$0.01 & 	0.33$\pm$0.03 & 	0.40$\pm$0.04 \\ 
ClG J0647+7015\dotfill & 		5.02$\pm$0.13 &	\nodata &	7.15$\pm$0.28 &	20.11$\pm$0.78 &	0.43$\pm$0.01 & 	0.59$\pm$0.02 & 	\nodata \\
ClG J0542-4100\dotfill & 		1.33$\pm$0.05 &	1.95$\pm$0.08 &	2.08$\pm$0.21 &	7.58$\pm$0.75 &	0.34$\pm$0.02 & 	0.47$\pm$0.03 & 	0.58$\pm$0.05 \\ 
ClG J1419+5326\dotfill & 		1.40$\pm$0.04 &	\nodata &	1.28$\pm$0.20 &	3.78$\pm$0.59 &	0.23$\pm$0.01 & 	0.33$\pm$0.03 & 	0.43$\pm$0.05 \\ 
ClG J0744+3927\dotfill & 		5.49$\pm$0.09 &	\nodata &	5.03$\pm$0.36 &	15.80$\pm$1.13 &	0.47$\pm$0.02 & 	0.64$\pm$0.04 & 	0.79$\pm$0.06 \\ 
ClG J1221+4918\dotfill & 		1.47$\pm$0.06 &	2.18$\pm$0.07 &	2.27$\pm$0.24 &	8.27$\pm$0.89 &	0.37$\pm$0.02 & 	0.55$\pm$0.05 & 	\nodata \\
ClG J1113-2615\dotfill & 		0.64$\pm$0.02 &	\nodata &	0.91$\pm$0.18 &	2.64$\pm$0.52 &	0.20$\pm$0.01 & 	0.27$\pm$0.02 & 	0.35$\pm$0.03 \\ 
ClG 1137+6625\dotfill & 		1.88$\pm$0.07 &	\nodata &	2.12$\pm$0.28 &	6.17$\pm$0.82 &	0.29$\pm$0.01 & 	0.41$\pm$0.03 & 	0.53$\pm$0.05 \\ 
RX J1350.0+6007\dotfill & 		0.60$\pm$0.05 &	1.04$\pm$0.05 &	1.01$\pm$0.20 &	4.05$\pm$0.80 &	0.26$\pm$0.02 & 	0.41$\pm$0.04 & 	0.53$\pm$0.07 \\ 
RX J1317+2911\dotfill & 		0.20$\pm$0.02 &	\nodata &	0.52$\pm$0.11 &	1.77$\pm$0.38 &	\nodata &	0.21$\pm$0.01 & 	0.32$\pm$0.03 \\ 
RX J1716+6708\dotfill & 		1.64$\pm$0.05 &	\nodata &	2.25$\pm$0.29 &	6.77$\pm$0.86 &	0.31$\pm$0.02 & 	0.43$\pm$0.03 & 	0.54$\pm$0.05 \\ 
ClG J1056-0337\dotfill & 		3.23$\pm$0.23 &	\nodata &	4.53$\pm$0.41 &	16.57$\pm$1.49 &	0.51$\pm$0.03 & 	0.62$\pm$0.04 & 	0.72$\pm$0.06 \\ 
ClG J1226+3332\dotfill & 		4.80$\pm$0.11 &	\nodata &	5.94$\pm$0.44 &	15.86$\pm$1.16 &	0.44$\pm$0.02 & 	0.59$\pm$0.03 & 	0.71$\pm$0.04 \\ 
ClG J1415+3611\dotfill & 		1.48$\pm$0.03 &	\nodata &	2.20$\pm$0.33 &	6.33$\pm$0.96 &	0.29$\pm$0.01 & 	0.40$\pm$0.02 & 	0.50$\pm$0.04 \\ 
ClG J1252-2927\dotfill & 		0.63$\pm$0.04 &	\nodata &	1.28$\pm$0.28 &	4.63$\pm$1.01 &	0.28$\pm$0.01 & 	0.42$\pm$0.03 & 	0.54$\pm$0.05 \\ 
\enddata
\label{tab:SSinfo}
\end{deluxetable*}

\tabletypesize{\scriptsize}
\begin{deluxetable*}{p{2.5cm}ccc}

\tablewidth{0pt}
\tabletypesize{\tiny}
\tablecaption{Cluster Measurements Assuming No Evolution}
\tablehead{
Cluster &
\colhead{$L_{\rm X,2500}$} &
\colhead{$M_{\rm g,2500}$} & \colhead{$M_{\rm g,500}$} \cr
& \colhead{($10^{44} L_\odot$)} &
\colhead{($10^{13} M_\odot$)} & \colhead{($10^{13} M_\odot$)}
}
\startdata
A665\dotfill & 	3.78$\pm$0.04 &	4.13$\pm$0.09 &	13.42$\pm$0.31 \\ 
A963\dotfill & 	\nodata &	2.99$\pm$0.09 &	9.63$\pm$0.28 \\ 
RX J0439.0+0520\dotfill & 	2.45$\pm$0.03 &	1.46$\pm$0.07 &	3.84$\pm$0.19 \\ 
A1423\dotfill & 	2.33$\pm$0.05 &	2.22$\pm$0.08 &	8.53$\pm$0.30 \\ 
ZwCl 2701\dotfill & 	2.38$\pm$0.02 &	1.64$\pm$0.07 &	4.76$\pm$0.21 \\ 
A773\dotfill & 	3.40$\pm$0.04 &	3.74$\pm$0.10 &	11.72$\pm$0.31 \\ 
A2261\dotfill & 	5.41$\pm$0.03 &	3.99$\pm$0.12 &	12.70$\pm$0.37 \\ 
ACO 2246\dotfill & 	\nodata &	0.59$\pm$0.05 &	2.01$\pm$0.16 \\ 
A1682\dotfill & 	1.95$\pm$0.08 &	2.17$\pm$0.09 &	7.94$\pm$0.33 \\ 
A2111\dotfill & 	2.11$\pm$0.06 &	2.78$\pm$0.09 &	9.43$\pm$0.30 \\ 
A267\dotfill & 	2.75$\pm$0.03 &	2.86$\pm$0.09 &	8.38$\pm$0.27 \\ 
RX J2129.7+0005\dotfill & 	4.91$\pm$0.03 &	2.87$\pm$0.12 &	8.37$\pm$0.34 \\ 
RX J0439.0+0715\dotfill & 	3.83$\pm$0.04 &	3.35$\pm$0.11 &	9.71$\pm$0.32 \\ 
A521\dotfill & 	2.58$\pm$0.09 &	2.71$\pm$0.11 &	11.42$\pm$0.47 \\ 
A1835\dotfill & 	12.13$\pm$0.06 &	5.50$\pm$0.17 &	13.62$\pm$0.42 \\ 
A68\dotfill & 	3.33$\pm$0.06 &	3.85$\pm$0.11 &	10.04$\pm$0.30 \\ 
MS 1455.0+2232\dotfill & 	5.74$\pm$0.01 &	2.27$\pm$0.13 &	6.20$\pm$0.34 \\ 
MS 1006.0+1202\dotfill & 	2.23$\pm$0.04 &	2.52$\pm$0.11 &	7.38$\pm$0.31 \\ 
A697\dotfill & 	6.48$\pm$0.08 &	6.38$\pm$0.17 &	19.24$\pm$0.52 \\ 
A611\dotfill & 	3.46$\pm$0.05 &	3.45$\pm$0.11 &	9.84$\pm$0.32 \\ 
ZwCl 3146\dotfill & 	10.85$\pm$0.02 &	4.32$\pm$0.19 &	10.88$\pm$0.49 \\ 
A781\dotfill & 	1.97$\pm$0.11 &	2.39$\pm$0.13 &	8.03$\pm$0.45 \\ 
MS 1008.1-1224\dotfill & 	2.41$\pm$0.04 &	2.61$\pm$0.12 &	7.69$\pm$0.36 \\ 
RXC J2245.0+2637\dotfill & 	4.16$\pm$0.03 &	2.65$\pm$0.14 &	6.94$\pm$0.36 \\ 
A1300\dotfill & 	5.09$\pm$0.09 &	4.71$\pm$0.16 &	17.25$\pm$0.60 \\ 
A2744\dotfill & 	\nodata &	6.67$\pm$0.20 &	16.77$\pm$0.51 \\ 
MS 2137.3-2353\dotfill & 	6.15$\pm$0.03 &	2.33$\pm$0.15 &	5.98$\pm$0.37 \\ 
A1995\dotfill & 	3.51$\pm$0.05 &	3.59$\pm$0.14 &	8.33$\pm$0.33 \\ 
ZwCl 1358+6245\dotfill & 	3.25$\pm$0.05 &	3.47$\pm$0.12 &	9.46$\pm$0.34 \\ 
A1722\dotfill & 	2.22$\pm$0.05 &	3.10$\pm$0.11 &	8.84$\pm$0.32 \\ 
RXC J0404.6+1109\dotfill & 	1.69$\pm$0.11 &	2.01$\pm$0.13 &	8.59$\pm$0.54 \\ 
RX J1532.9+3021\dotfill & 	10.90$\pm$0.04 &	3.78$\pm$0.23 &	10.15$\pm$0.61 \\ 
A370\dotfill & 	3.49$\pm$0.07 &	4.25$\pm$0.18 &	11.62$\pm$0.50 \\ 
ZwCl 1953\dotfill & 	4.48$\pm$0.06 &	4.02$\pm$0.20 &	11.78$\pm$0.58 \\ 
RXC J0949.8+1707\dotfill & 	5.30$\pm$0.08 &	4.42$\pm$0.22 &	13.00$\pm$0.63 \\ 
ClG J1416+4446\dotfill & 	1.48$\pm$0.03 &	1.27$\pm$0.13 &	4.06$\pm$0.43 \\ 
RXC J2228.6+2036\dotfill & 	5.71$\pm$0.11 &	4.92$\pm$0.25 &	14.68$\pm$0.75 \\ 
MS 0302.7+1658\dotfill & 	\nodata &	1.27$\pm$0.15 &	4.09$\pm$0.49 \\ 
MS 1621.5+2640\dotfill & 	1.98$\pm$0.06 &	2.50$\pm$0.17 &	8.13$\pm$0.54 \\ 
MACS J0417.5-1154\dotfill & 	14.27$\pm$0.22 &	8.17$\pm$0.38 &	25.86$\pm$1.21 \\ 
RXC J1206.2-0848\dotfill & 	10.79$\pm$0.12 &	7.91$\pm$0.30 &	23.15$\pm$0.89 \\ 
ClG J0329-0212\dotfill & 	6.94$\pm$0.04 &	3.19$\pm$0.24 &	10.32$\pm$0.79 \\ 
RX J1347.5-1145\dotfill & 	24.62$\pm$0.06 &	10.58$\pm$0.35 &	26.60$\pm$0.89 \\ 
ClG J1701+6414\dotfill & 	1.71$\pm$0.03 &	1.64$\pm$0.16 &	5.69$\pm$0.55 \\ 
3C 295\dotfill & 	3.86$\pm$0.02 &	2.19$\pm$0.18 &	5.85$\pm$0.47 \\ 
ClG J1621+3810\dotfill & 	4.17$\pm$0.04 &	2.92$\pm$0.20 &	8.52$\pm$0.58 \\ 
ClG J1524+0957\dotfill & 	1.48$\pm$0.05 &	1.83$\pm$0.20 &	5.43$\pm$0.59 \\ 
MS 0451.6-0305\dotfill & 	8.82$\pm$0.11 &	6.74$\pm$0.37 &	15.49$\pm$0.86 \\ 
MS 0015.9+1609\dotfill & 	8.78$\pm$0.10 &	7.14$\pm$0.40 &	20.52$\pm$1.15 \\ 
ClG J1149+2223\dotfill & 	7.55$\pm$0.18 &	6.86$\pm$0.38 &	22.15$\pm$1.23 \\ 
ClG J1423+2404\dotfill & 	7.87$\pm$0.05 &	3.03$\pm$0.31 &	8.36$\pm$0.84 \\ 
ClG J1354-0221\dotfill & 	0.85$\pm$0.05 &	1.19$\pm$0.16 &	3.87$\pm$0.52 \\ 
ClG J0717+3745\dotfill & 	12.73$\pm$0.23 &	10.46$\pm$0.50 &	31.03$\pm$1.48 \\ 
ClG J1120+2326\dotfill & 	1.35$\pm$0.05 &	1.60$\pm$0.21 &	4.17$\pm$0.55 \\ 
ClG J2129-0741\dotfill & 	6.34$\pm$0.21 &	6.23$\pm$0.30 &	17.66$\pm$0.86 \\ 
MS 2053.7-0449\dotfill & 	1.37$\pm$0.03 &	1.31$\pm$0.19 &	3.88$\pm$0.57 \\ 
ClG J0647+7015\dotfill & 	7.34$\pm$0.16 &	7.89$\pm$0.31 &	21.30$\pm$0.83 \\ 
ClG J0542-4100\dotfill & 	2.29$\pm$0.08 &	2.59$\pm$0.26 &	8.68$\pm$0.86 \\ 
ClG J1419+5326\dotfill & 	2.13$\pm$0.05 &	1.44$\pm$0.23 &	4.12$\pm$0.65 \\ 
ClG J0744+3927\dotfill & 	8.90$\pm$0.13 &	5.86$\pm$0.42 &	18.02$\pm$1.29 \\ 
ClG J1221+4918\dotfill & 	2.69$\pm$0.07 &	2.97$\pm$0.32 &	9.09$\pm$0.97 \\ 
ClG J1113-2615\dotfill & 	1.02$\pm$0.04 &	1.06$\pm$0.21 &	2.73$\pm$0.54 \\ 
ClG 1137+6625\dotfill & 	3.23$\pm$0.10 &	2.46$\pm$0.33 &	6.65$\pm$0.89 \\ 
RX J1350.0+6007\dotfill & 	1.31$\pm$0.07 &	1.41$\pm$0.28 &	4.90$\pm$0.97 \\ 
RX J1317+2911\dotfill & 	0.38$\pm$0.04 &	0.68$\pm$0.15 &	1.87$\pm$0.40 \\ 
RX J1716+6708\dotfill & 	2.96$\pm$0.08 &	2.70$\pm$0.34 &	7.26$\pm$0.92 \\ 
ClG J1056-0337\dotfill & 	6.67$\pm$0.16 &	6.14$\pm$0.55 &	19.02$\pm$1.71 \\ 
ClG J1226+3332\dotfill & 	8.67$\pm$0.17 &	6.71$\pm$0.49 &	16.52$\pm$1.21 \\ 
ClG J1415+3611\dotfill & 	3.03$\pm$0.08 &	2.68$\pm$0.41 &	6.53$\pm$0.99 \\ 
ClG J1252-2927\dotfill & 	\nodata &	1.90$\pm$0.41 &	6.25$\pm$1.36 \\ 
\enddata
\label{tab:noEinfo}
\end{deluxetable*}

\end{document}